\documentclass[lettersize,journal]{IEEEtran}
\usepackage{stmaryrd}
\usepackage{amsmath}
\usepackage{amssymb}
\usepackage{psfrag}
\usepackage{graphicx}
\usepackage{epstopdf}
\usepackage{todonotes}
\usepackage{multirow}
\usepackage{booktabs}
\usepackage{siunitx}
\usepackage{cite}
\usepackage{algorithm}
\usepackage{svg}
\usepackage{mathtools}
\usepackage[bookmarks=false]{hyperref}
\usepackage{tabularx}
\usepackage{algpseudocode}
\usepackage{float}
\DeclareMathOperator*{\argmax}{argmax}
\ifCLASSOPTIONcompsoc
\usepackage[caption=false,font=normalsize,labelfont=sf,textfont=sf]{subfig}
\else
\usepackage[caption=false,font=footnotesize]{subfig}
\fi

\makeatletter
\newcommand{\multiline}[1]{%
  \begin{tabularx}{\dimexpr\linewidth-\ALG@thistlm}[t]{@{}X@{}}
    #1
  \end{tabularx}
}
\newcommand{\commentline}[1]{
    \State \textbf{\# #1}
}
\makeatother

\begin{document}
%
\title{Detect to Learn: Structure Learning with Attention and Decision Feedback for MIMO-OFDM Receive Processing}

\author{Jiarui Xu, Lianjun Li, Lizhong Zheng, and Lingjia Liu
\thanks{J. Xu, L. Li and L. Liu are with Wireless@Virginia Tech, Department of Electrical \& Computer Engineering, Virginia Tech, Blacksburg, Virginia 24061. L. Zheng is with the Department of Electrical Engineering and Computer Science, Massachusetts Institute of Technology, Cambridge, MA 02139.
The work is supported by US National Science Foundation (NSF) and Intel under grants CNS-2003059, CNS-2002908 and CCF-1937487.}
}
%



\maketitle

\begin{abstract}
The limited over-the-air (OTA) pilot symbols in multiple-input-multiple-output orthogonal-frequency-division-multiplexing (MIMO-OFDM) systems presents a major challenge for detecting transmitted data symbols at the receiver, especially for machine learning-based approaches. 
While it is crucial to explore effective ways to exploit pilots, one can also take advantage of the data symbols to improve detection performance.
Thus, this paper introduces an online attention-based approach, namely RC-AttStructNet-DF, that can efficiently utilize pilot symbols and be dynamically updated with the detected payload data using the decision feedback (DF) mechanism. 
Reservoir computing (RC) is employed in the time domain network to facilitate efficient online training. 
The frequency domain network adopts the novel 2D multi-head attention (MHA) module to capture the time and frequency correlations, and the structural-based StructNet to facilitate the DF mechanism. 
The attention loss is designed to learn the frequency domain network. 
The DF mechanism further enhances detection performance by dynamically tracking the channel changes through detected data symbols. 
The effectiveness of the RC-AttStructNet-DF approach is demonstrated through extensive experiments in MIMO-OFDM and massive MIMO-OFDM systems with different modulation orders and under various scenarios.
\end{abstract}


%
\IEEEpeerreviewmaketitle


\section{Introduction}
Deep learning (DL) has attracted significant attention due to its overwhelming privilege in computer vision (CV), natural language processing (NLP), and robotics.
DL has also demonstrated great potential in MIMO-OFDM symbol detection tasks.
Efforts have been devoted to utilizing multi-layer perceptron (MLP)~\cite{Ye2018MIMO}, convolution neural network (CNN)~\cite{zhao2021deep, chen2019efficient, honkala2021deeprx}, long short-term memory (LSTM) network~\cite{Lyu2020DeepNN, Liao2019DeepNN}, or generative adversarial network (GAN)~\cite{Yi2020JointEst} to jointly estimate the channel and detect symbols.
These approaches do not rely on explicit modeling of the system and the estimation of the channel state information (CSI), but instead, they treat the model as a black box and use neural networks (NNs) to deal with more complicated wireless systems with non-linear device components.
Moreover, DL-based approaches offer a new avenue for symbol detection tasks with a large number of OFDM subcarriers and high modulation orders, thanks to their lower computational complexity compared to conventional optimal detection algorithms~\cite{shafin2020artificial}.


While DL-based approaches are promising, such approaches have their own challenges to be applied in real-world MIMO-OFDM systems.
One of the main challenges is the data requirement.
Training NNs requires a huge volume of labeled data along with a long training time.
In modern cellular networks, OTA labeled data is scarce and extremely expensive to obtain, resulting in limited training data for symbol detection tasks.
Attempts have been made to address such issues by training with extensive channel realizations offline and initializing the network with offline weights for online adaptation~\cite{Ye2018MIMO}.
However, such approaches are hard to be adopted to the dynamically changing environments in practice, where the systems have different transmission modes with link adaptation, rank adaptation, and schedules operating on a subframe (1 millisecond) basis~\cite{shafin2020artificial, 4GMIMO_OFDM}.
The discrepancy between modes for offline and online training can result in model mismatch and prohibit the offline weights from being utilized online.

Another challenge is how to design an interpretable network with domain knowledge.
Learning both the known and unknown features requires a large amount of training data.
The desired network should embed what we have known into the architecture instead of learning everything from scratch.
Recent advances employ deep-unfolding NNs to improve network interoperability for detection~\cite{samuel2019learning, he2018model, khani2020adaptive, he2020oamp2}, transceiver design~\cite{shi2022deep, kang2022mixed}, and channel estimation~\cite{he2018deep}.
Specifically, for symbol detection, such approaches~\cite {samuel2019learning, he2020oamp2, he2018model, khani2020adaptive} build the network by unfolding conventional iterative algorithms and thus reduce the number of trainable parameters in the network.
For example, MMNet~\cite{khani2020adaptive} converts each iteration of the iterative soft-thresholding algorithms into network layers and only learns a few parameters of the algorithm with NN in each layer.
It has shown superior performance in independent and identically distributed (i.i.d.) Gaussian channels and 3D MIMO channels released by 3rd Generation Partnership Project (3GPP)~\cite{study3d3gpp}.
However, due to the stack of multiple layers, the number of parameters grows, leading to a demand for a large-size training dataset.
Furthermore, the assumption of the perfect CSI knowledge also limits its utilization in practice.

Our previous RC based approaches~\cite{mosleh2017brain, zhou2019, zhou2020rcnet} adopts the RC in the time domain to conduct symbol detection and thus can be learned with only a limited amount of online training data.
More recently, RC-Struct~\cite{xu2021rcstruct} extends the idea of the time domain RC-based approaches by introducing an extra frequency domain structure-based network.
The network consists of a time domain RC and a frequency domain classifier.
By processing in both domains, the time domain convolution and superposition operation of the wireless channel, the time-frequency structure of OFDM, and the frequency domain structure of the repetitive modulation constellation pattern, can all be embedded into the underlying NNs to take advantage of the available structural knowledge of the MIMO-OFDM system.
The incorporation of such structural knowledge helps to significantly reduce the needed training overhead and improve the system performance in a realistic environment.
However, these RC-based methods only learn from the pilot OFDM symbols and then test on the rest data OFDM symbols. 
While they still work in the scenario where the channel is gradually evolving over different OFDM symbols within a subframe, it does not learn the dynamic features that are provided by the data symbols.

In this paper, we introduce an online attention-based approach, RC-AttStructNet-DF, to tackle the above challenges.
We follow our previous work, RC-Struct, to embed structural knowledge of the MIMO-OFDM system into the network architecture. 
To further improve detection performance, we adopt an additional attention mechanism in our approach, which is achieved by a 2D MHA module and attention loss.
Inspired by the MHA module in the Transformer~\cite{vaswani2017attention}, we develop the 2D MHA to capture time and frequency correlations in a two-dimensional manner.
The attention loss is designed for the frequency domain network to assign different weights to the training loss of different training samples based on their confidence levels. 
Moreover, unlike RC-Struct which only learns from the pilot symbols, RC-AttStructNet-DF also takes advantage of the data symbols.
Specifically, a DF mechanism, which takes the detected data symbols as the training data, is used to dynamically track the channel variations within a subframe.
To mitigate the error propagation of the DF procedure, the frequency domain StructNet is designed on top of the RC-Struct, where a parameter estimation (PE) layer is additionally adopted.
The customized StructNet is shown to be robust to incorrect labels owing to the embedded structural information, making it a suitable network to be applied with the DF process. 
The introduced PE layer further facilitates the DF mechanism by learning the network parameters with the underlying NN.
Evaluation results demonstrate the effectiveness of adopting the attention mechanism to improve detection performance, and the benefits of employing the StructNet along with the DF mechanism to mitigate the issue of error propagation and dynamically update network parameters. 
Meanwhile, extensive experiments have been conducted to show the considerable performance gain of RC-AttStructNet-DF over the conventional model-based methods and the state-of-the-art learning-based approaches in both the MIMO-OFDM system and the massive MIMO-OFDM system under the 3GPP 3D channel~\cite{study3d3gpp}.
Evaluation results also demonstrate that RC-AttStructNet-DF can perform better than the conventional approaches with a practical pilot pattern specified by the 3GPP 5G NR.
Our contributions are summarized as the following:
\begin{itemize}
    \item 
    We develop an attention-based symbol detection approach, where attention is achieved by adopting a novel 2D MHA module and attention loss.
    The 2D MHA is constructed based on the MHA module in the Transformer, which captures time and frequency correlations in a two-dimensional manner.
    The attention loss is designed to assign different weights to the training loss of different training samples according to their confidence levels. 
    \item 
    We exploit the DF mechanism to take advantage of the data symbols to improve detection performance. 
    Specifically, the DF algorithm updates the network weights by utilizing the detected data symbols as the training data so that the channel changes across different OFDM data symbols can be dynamically tracked.
    \item We design a learning-efficient network, StructNet, in the frequency domain on top of the existing RC-Struct. The customized StructNet is shown to be robust to incorrect labels owing to the embedded structural information, which effectively mitigates the issue of error propagation typically encountered in the DF procedure. 
    Moreover, the introduced PE layer in StructNet enables the network to dynamically update its parameters, further facilitating the underlying DF mechanism.
    \item Extensive performance evaluations have been conducted to show that RC-AttStructNet-DF outperforms RC-Struct in the MIMO-OFDM system under various scenarios with relatively high user mobility. 
    Furthermore, RC-AttStructNet-DF demonstrates its effectiveness in large-scale antenna array settings, extending the application scenario from the MIMO system to the massive MIMO system.
    Experimental results also reveal the potential of employing RC-AttStructNet-DF in a practical pilot pattern specified by the 3GPP 5G NR.
\end{itemize}

The rest of the paper is organized as follows.
Sec.~\ref{sec:problem_statement} presents the MIMO-OFDM system and problem statement.
Sec.~\ref{sec:preliminary} briefly discusses RC, the MHA module, and the atomic decision neuron network (ADNN) exploited in RC-Struct.
Sec.~\ref{sec:structnet_analysis} introduces the frequency-domain network, StructNet, in a simple MIMO system. 
Then Sec.~\ref{sec:introduced_method} discusses how to apply StructNet in the MIMO-OFDM system and introduces the RC-AttStructNet-DF.
Complexity analysis is provided in Sec.~\ref{sec:complexity_analysis}.
A toy experiment for analyzing StructNet is shown in Sec.~\ref{sec:toy_experiment_mimo}.
Sec.~\ref{sec:evaluations_mimo_ofdm} compares the performance of RC-AttStructNet-DF with state-of-the-art methods in both the MIMO-OFDM system and the massive MIMO-OFDM system.
The paper is concluded in Sec.~\ref{sec:conclusion}.

\textbf{Notations}: Scalar, vector, and matrix are denoted by non-bold letter, bold lowercase letter, and bold uppercase letter.
The superscripts $(\cdot)^t$ and $(\cdot)^f$ are used to differentiate the time domain and the frequency domain signals, respectively.
$\mathbb{C} (\mathbb{R})$ represents the complex (real) number set.
Other sets are denoted by calligraphic letters.

\section{Problem Formulation}
\label{sec:problem_statement}



We consider a MIMO-OFDM system with $N_t$ transmit antennas, $N_r$ receive antennas, and $N_{\mathrm{sc}}$ subcarriers.
The information is modulated in the frequency domain on a subframe basis, where each subframe consists of $N$ OFDM symbols.
The $n$-th OFDM symbol in the frequency domain can be represented as $\boldsymbol{X}_n^f \in {\mathbb C}^{N_t \times N_{\mathrm{sc}}}$.
The frequency domain OFDM symbols are converted to the time domain through an inverse fast Fourier transform (IFFT) and cyclic prefix (CP) addition with length $N_{\mathrm{cp}}$.
Then all the time domain OFDM symbols are concatenated together to form the transmitted time domain signal $\boldsymbol{X}^t  = \left[ \boldsymbol{X}_0^t, \boldsymbol{X}_1^t, \dots, \boldsymbol{X}_{N-1}^t \right] \in \mathbb{C}^{N_t \times N(N_{\mathrm{sc}} + N_{\mathrm{cp}})}$, where $\boldsymbol{X}_n^t \in \mathbb{C}^{N_t \times (N_{\mathrm{sc}} + N_{\mathrm{cp}})}$ is the $n$-th OFDM symbol in the time domain.

\begin{table}[!t]
\centering
\caption{Notation in the system}
\resizebox{\columnwidth}{!}{
\begin{tabular}{ll}
\toprule
\textbf{Symbol} & \textbf{Definition} \\
\midrule
$N_t$ & Number of transmit antennas \\
$N_r$ & Number of receive antennas \\
$N_{\mathrm{sc}}$ & Number of OFDM subcarriers \\
$N_{\mathrm{cp}}$ & Length of cyclic prefix \\
$N_p$ & Number of pilot symbols in one OFDM frame (training set) \\
$N_d$ & Number of data symbols in one OFDM frame (testing set) \\ 
$N$ & Total number of symbols in one OFDM frames \\ 
$L$ & Total number of channel delays \\
$\boldsymbol{Y}^t \in {\mathbb C}^{N_r\times N(N_{\mathrm{sc}} + N_{\mathrm{cp}})}$ & The received time domain signal \\
$\boldsymbol{X}^t \in {\mathbb C}^{N_t\times N(N_{\mathrm{sc}} + N_{\mathrm{cp}})}$ & The transmitted time domain signal \\
$\boldsymbol{h}_{n_r, n_t}^t \in {\mathbb C}^{L}$  & The channel impulse response between receiver $n_r$ and transmitter $n_t$ \\
$\boldsymbol{y}_{n_r}^t \in {\mathbb C}^{N(N_{\mathrm{sc}} + N_{\mathrm{cp}})}$ & The received signal at $n_r$-th receiver in time domain \\
$\boldsymbol{x}_{n_t}^t \in {\mathbb C}^{N(N_{\mathrm{sc}} + N_{\mathrm{cp}}})$ & The transmitted signal at $n_t$-th transmitter in time domain \\
$\boldsymbol{Y}_n^t \in {\mathbb C}^{N_r\times (N_{\mathrm{sc}} + N_{\mathrm{cp}})}$ & The received $n$-th OFDM symbols in time domain \\
$\boldsymbol{X}_n^t \in {\mathbb C}^{N_t\times (N_{\mathrm{sc}} + N_{\mathrm{cp}})}$ & The transmitted $n$-th OFDM symbols in time domain \\
$\boldsymbol{Y}_n^f \in {\mathbb C}^{N_r\times N_{\mathrm{sc}}}$ & The received $n$-th OFDM symbols in frequency domain \\
$\boldsymbol{X}_n^f \in {\mathbb C}^{N_t\times N_{\mathrm{sc}}}$ & The transmitted $n$-th OFDM symbols in frequency domain \\
\bottomrule
\end{tabular}
}
\label{tab:notations}
\end{table}
 
Denote the time domain signal at the $n_t$-th transmit antenna as $\boldsymbol{x}_{n_t}^t \in \mathbb{C}^{N(N_{\mathrm{sc}} + N_{\mathrm{cp}})}$, and the $L$-tap channel between the receive antenna $n_r$ and transmit antenna $n_t$ as $\boldsymbol{h}_{n_r, n_t}^t \in {\mathbb C}^{L}$, where the channel is gradually evolving across different OFDM symbols.
Then at the receiver side, the received signal can be written as
\begin{align}
    \boldsymbol{y}_{n_r}^t = \sum_{n_t=0}^{N_t-1} \boldsymbol{h}_{n_r, n_t}^t \circledast g(\boldsymbol{x}_{n_t}^t) + \boldsymbol{w}^t,
\end{align}
where $\boldsymbol{y}_{n_r}^t \in \mathbb{C}^{N(N_{\mathrm{sc}} + N_{\mathrm{cp}})}$ is the received signal at the $n_r$-th receive antenna; $\circledast$ stands for the convolution operation; $g(\cdot)$ represents the non-linear distortion caused by transmitter components, such as power amplifier (PA); $\boldsymbol{w}^t$ denotes the additive white Gaussian noise (AWGN).
Let $\boldsymbol{Y}^t \in \mathbb{C}^{N_r \times N(N_{\mathrm{sc}} + N_{\mathrm{cp}})}$ represent the received time domain signal for all the receive antennas, and $\boldsymbol{Y}_n^t \in \mathbb{C}^{N_r \times (N_{\mathrm{sc}} + N_{\mathrm{cp}})}$ is the $n$-th received OFDM symbol in the time domain.
The frequency domain counterpart of the received signal $\boldsymbol{Y}_n^f \in {\mathbb C}^{N_r \times 
 N_{\mathrm{sc}}}$ is acquired by removing CP and following with a fast Fourier transform (FFT).
The notations are summarized in Table~\ref{tab:notations}.



The MIMO-OFDM symbol detection task is to recover the transmitted symbol $\boldsymbol{X}_n^f$ from the received time-domain observations $\boldsymbol{Y}_n^t$ for each OFDM symbol.
Pilot symbols, which are known at both the transmitter and receiver sides, will be embedded in each subframe to facilitate the detection of unknown data symbols.
For ease of discussion, we assume the first $N_p$ OFDM symbols are the pilots, and the rest of the $N_d$ symbols are the unknown data.
However, the introduced approach can also be applied to other pilot patterns, which will be shown in Sec.~\ref{exp:scattered_pilot_pattern_exp}.
In this work, the training dataset for our NN-based approach consists of the transmitted and received time domain pilot signals $\boldsymbol{X}_n^t$ and $\boldsymbol{Y}_n^t$ along with the frequency domain pilot symbols $\boldsymbol{X}_n^f$.

\section{Preliminary}
\label{sec:preliminary}
\subsection{Reservoir computing}
RC is a recurrent neural network (RNN) based approach that can process sequential data. 
It only conducts training on the output layer and fixes the randomly initialized weights of the input layer and the RNN-based reservoir.
As the detailed discussion about RC has been provided in our previous work~\cite{mosleh2017brain, zhou2019, zhou2020deep, zhou2020rcnet, li2020reservoir, xu2021rcstruct}, we mainly focus on the processing procedures of RC.
Denote the input data as $\boldsymbol{U} \in \mathbb{C}^{N_{in} \times T}$, where $\boldsymbol{u}(m)$ represents the $m$-th column of $\boldsymbol{U}$.
The processing of RC at the $m$-th time step can be represented by the following state transition equation and the output equation:
\begin{gather}
    \boldsymbol{s}(m) = f(\boldsymbol{W}\boldsymbol{s}(m-1) + \boldsymbol{W}_{\mathrm{in}} \; \boldsymbol{u}(m)), \;
    \boldsymbol{\hat{o}}(m) = \boldsymbol{W}_{\mathrm{out}} \; \boldsymbol{z}(m),\label{equ:rc_output}
\end{gather}
where $\boldsymbol{s}(m) \in \mathbb{C}^{N_n}$ is the state vector; $\boldsymbol{W}_{\mathrm{in}} \in \mathbb{C}^{N_n \times N_{in}}$ and $\boldsymbol{W} \in \mathbb{C}^{N_n \times N_n}$ are the fixed input weight matrix and reservoir transition matrix, respectively; the $f(\cdot)$ is the hyperbolic tangent function; $\boldsymbol{\hat{o}}(m) \in \mathbb{C}^{N_{\mathrm{out}}}$ is the predicted output at $m$-th time step; $\boldsymbol{z}(m) = \left[ \boldsymbol{s}(m)^T, \boldsymbol{u}(m)^T \right]^T \in \mathbb{C}^{N_n+N_{in}}$ is the state and input concatenation; $\boldsymbol{W}_{\mathrm{out}} \in \mathbb{C}^{N_{\mathrm{out}} \times (N_n + N_{in})}$ is the trainable output weight matrix.
During training, the output weight matrix is learned by the least square (LS) solution
\begin{align}
    \label{equ:rc_least_square}
    \boldsymbol{\hat{W}}_{\mathrm{out}} = \arg\min_{\boldsymbol{W}_{\mathrm{out}}} \| \boldsymbol{W}_{\mathrm{out}}\boldsymbol{Z} - \boldsymbol{O} \|_F^2 = \boldsymbol{O}\boldsymbol{Z}^\dag,
\end{align}
where $\boldsymbol{Z} = [\boldsymbol{z}(0), \boldsymbol{z}(1), \dots, \boldsymbol{z}(T-1)] \in \mathbb{C}^{(N_n+N_{in})\times T}$ is the concatenation of $\boldsymbol{z}(m)$, $\boldsymbol{O} \in \mathbb{C}^{N_{\mathrm{out}}\times T}$ is the training label, and $\boldsymbol{Z}^\dag$ is the Moore-Penrose inverse of $\boldsymbol{Z}$.



\subsection{Multi-head attention}

\begin{figure}
\centering
\includegraphics[width=0.35\linewidth]{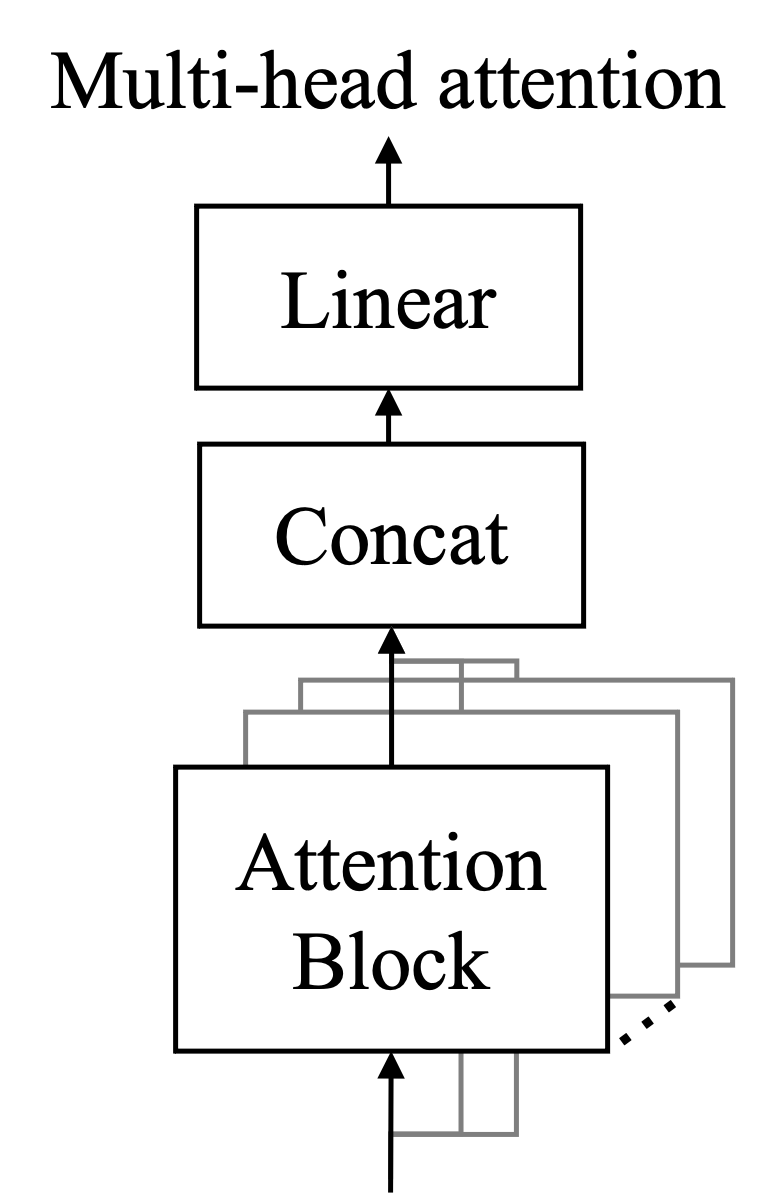}
\caption{The MHA module.}
\label{figs:multi_head_attention}
\vspace{-1em}
\end{figure}

The MHA module is first introduced in the Transformer~\cite{vaswani2017attention}, which allows the model to attend to information from different representation subspaces.
As shown in Fig.~\ref{figs:multi_head_attention}, the MHA module consists of multiple attention blocks.
For each block, the attention function maps the input into the query, key, and value subspaces as the query, key, and value embeddings.
The output of the attention block, or the attention head, is obtained by the scaled dot product of the three embeddings.
Particularly, the attention block can be computed by the following equations:
\begin{gather}
    \boldsymbol{Q} = \boldsymbol{U}\cdot\boldsymbol{W}_q,
    \boldsymbol{K} = \boldsymbol{U}\cdot\boldsymbol{W}_k,
    \boldsymbol{V} = \boldsymbol{U}\cdot\boldsymbol{W}_v,\\
    \mathrm{head} = f_{sm}(\frac{\boldsymbol{Q}\boldsymbol{K}^T}{\sqrt{N_k}})\cdot\boldsymbol{V},
\end{gather}
where $\boldsymbol{U} \in \mathbb{R}^{N_{in1} \times N_{in2}}$ is the input to the attention block; $\boldsymbol{W}_q \in \mathbb{R}^{N_{in2} \times N_{k}}$, $\boldsymbol{W}_k \in \mathbb{R}^{N_{in2} \times N_{k}}$, and $\boldsymbol{W}_v \in \mathbb{R}^{N_{in2} \times N_{v}}$ are linear projections that project the input to the query, key, and value embedding spaces, respectively; $f_{sm}(\cdot)$ represents the softmax function; $N_k$ and $N_v$ are the dimensions of key embedding and value embedding, respectively.
In the MHA module, the heads of each attention block are aggregated by concatenation and a linear projection:
\begin{align}
    \mathrm{MultiHead} = \mathrm{Concat}(\mathrm{head}_0, \mathrm{head}_1, \dots, \mathrm{head}_{N_{a} - 1})\cdot \boldsymbol{W}_o,
\end{align}
where $\boldsymbol{W}_o \in \mathbb{R}^{N_aN_{v} \times N_o}$ is the output linear project, $N_a$ is the number of attention blocks, and $N_o$ is the output dimension.

\subsection{Atomic decision neuron network}

The ADNN~\cite{zhou2020learning2, xu2021rcstruct} is a frequency domain symbol detection network that utilizes a single binary classifier to conduct the multi-class detection task.
The design of the network exploits the repetitive structure of the modulation constellation, which improves the training efficiency by reducing the network size and utilizing training samples more efficiently.

\subsubsection{Problem formulation}
Consider a $N_r \times N_t$ MIMO system in the frequency domain,
\begin{align}
    \boldsymbol{y}^f = \boldsymbol{H}^f\boldsymbol{x}^f + \boldsymbol{w}^f,
\end{align}
where $\boldsymbol{y}^f \in \mathbb{C}^{N_r}$ is the received signal, $\boldsymbol{H}^f \in \mathbb{C}^{N_r \times N_t}$ represents the channel in the frequency domain, $\boldsymbol{x}^f \in \mathbb{C}^{N_t}$ stands for the transmitted $M$ quadrature amplitude modulation ($M$-QAM) symbol, and $\boldsymbol{w}^f \in \mathbb{C}^{N_r}$ is the additive noise that may not need to be Gaussian noise.
The real-valued form of $\boldsymbol{y}^f$ and $\boldsymbol{x}^f$, which are used as the input and output of the network, are obtained by $\boldsymbol{\tilde{y}}^f = f_R(\boldsymbol{y}^f)$ and $\boldsymbol{\tilde{x}}^f = f_R(\boldsymbol{x}^f) \in \mathcal{A}^{2N_t}$, where $f_R(\cdot)$ is the complex to real value function $f_R(\boldsymbol{x}) = \left[ \Re{\{\boldsymbol{x}\}}^T, \Im{\{\boldsymbol{x}}\}^T \right]^T$ and $\mathcal{A}$ is the set $\{-2K-1, -2K+1, \dots, 2K-1, 2K+1\}$ with $K = \frac{\sqrt{M}-2}{2}$.
Note that the $\boldsymbol{\tilde{x}}^f$ are now the $\sqrt{M}$ pulse amplitude modulation ($\sqrt{M}$-PAM) symbols.
The real-valued form of channel is acquired by:
\begin{align}
    \boldsymbol{\Tilde{H}}^f = \begin{bmatrix}
    \Re{\{\boldsymbol{H}^f\}}, -\Im{\{\boldsymbol{H}^f\}}\\
    \Im{\{\boldsymbol{H}^f\}}, \Re{\{\boldsymbol{H}^f\}}
     \end{bmatrix}.
\end{align}

Then the problem becomes a classification task with labels in the set $\mathcal{A}$, which can be approximated by the Naive Bayesian approximation
\begin{align}
\label{eq:mimo_obj}
    \argmax_{\boldsymbol{\tilde{x}}^f} P\{\boldsymbol{\tilde{x}}^f | \boldsymbol{\tilde{y}}^f\} \approx \argmax_{\boldsymbol{\tilde{x}}^f} \prod_{n \in \{0, 1, \dots, 2N_t-1\}} P_n\{{\tilde{x}}_n^f | \boldsymbol{\tilde{y}}^f \},
\end{align}
where the $P_n\{\cdot | \boldsymbol{\tilde{y}}^f\}$ represents the marginal distribution of the $n$-th entry of $\boldsymbol{\tilde{x}}^f$.
The ADNN is designed to learn functions that approximate the $P_n\{\cdot | \boldsymbol{\tilde{y}}^f\}$.


\subsubsection{Binary detection}
For simplicity, we start with the QPSK detection, where the ${\tilde{x}}_n^f$ is in the class set $\{+1, -1\}$.
Denote the function approximated by the binary classifier as $\mathcal{B}_n$.
We can directly apply the binary classifier to estimate the marginal likelihood ratio with
\begin{align}
    {P_n\{{\hat{x}}_n^f = +1 | \boldsymbol{\tilde{y}}^f\}\over
    P_n\{{\hat{x}}_n^f = -1 | \boldsymbol{\tilde{y}}^f\}} \approx {\mathcal{B}_n(\hat{b}_n = +1; \boldsymbol{\tilde{y}}^f)\over
    \mathcal{B}_n(\hat{b}_n = -1; \boldsymbol{\tilde{y}}^f)} =: {\mathcal L}_b^{+-}(\boldsymbol{\tilde{y}}^f),
\end{align}
where ${\hat{x}}_n^f$ is the estimated transmitted symbol, $\hat{b}_n$ is the predicted binary label, and ${\mathcal L}_b^{+-}$ denotes the likelihood ratio of the binary classifier.

\subsubsection{Multi-class detection}
\label{sec:prelim_adnn_multi_class}

\begin{figure}
\centering
\includegraphics[width=0.7\linewidth]{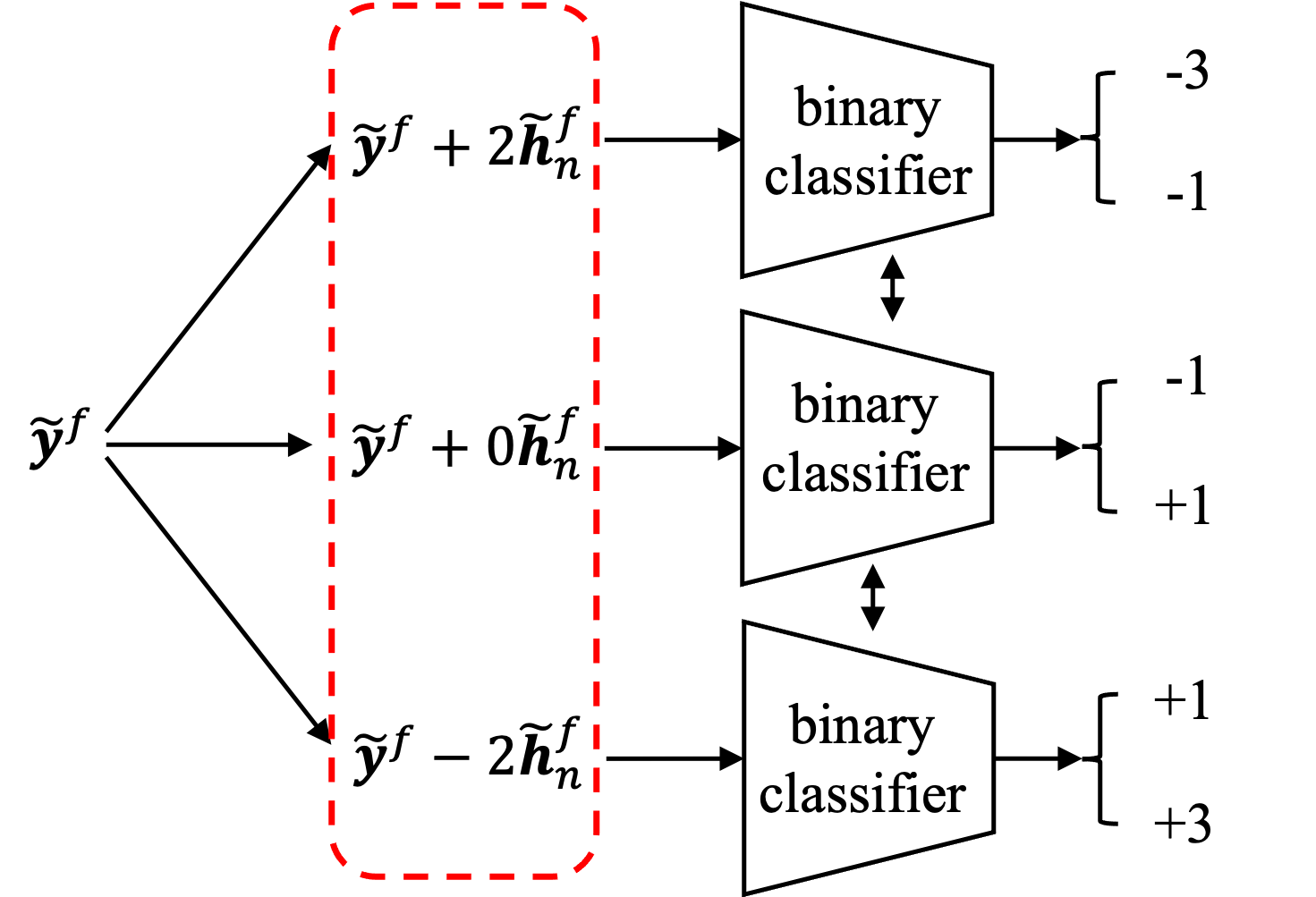}
\caption{
The architecture of ADNN. 
``Share" means that the weights of all the binary classifiers are shared.}
\vspace{-1em}
\label{figs:adnn_structure}
\end{figure}

When transmitting the $M$-QAM symbol, we perform the multi-class classification task with the class set $\mathcal{A}$ adopting the following shifting principle~\cite{zhou2020learning2}:
\begin{gather}
{{P_n\{{\hat{x}}_n^f = -2k+1| \boldsymbol{\tilde{y}}^f\}}\over
{P_n\{{\hat{x}}_n^f = -2k-1 | \boldsymbol{\tilde{y}}^f\}}} =
{P_n\{{\hat{x}}_n^f = +1 | \boldsymbol{\tilde{y}}^f+2k\cdot \boldsymbol{\tilde{h}}_n^f\}\over
    P_n\{{\hat{x}}_n^f = -1 | \boldsymbol{\tilde{y}}^f+2k\cdot \boldsymbol{\tilde{h}}_n^f\}},
\end{gather}
where $\boldsymbol{\tilde{h}}_n^f$ is the $n$-th column of the channel $\boldsymbol{\Tilde{H}}^f$ and $k =-K, -K+1, \dots, +K$.
The principle is based on the fact that the constellation points of the QAM modulation share the same distance $2$ and have a repetitive structure.
By shifting the received signal $\boldsymbol{\tilde{y}}^f$ in the direction of $\boldsymbol{\tilde{h}}_n^f$ with step size $2k$, which corresponds to $\boldsymbol{\tilde{y}}^f+2k\cdot \boldsymbol{\tilde{h}}_n^f$, the transmit symbol $-2k+1$ and $-2k-1$ are shifted to $+1$ and $-1$, respectively.
Note that the binary label $+1$ and $-1$ is determined by the step size plus the transmit symbol.
Then the likelihood ratio between class $-2k+1$ and $-2k-1$ can be estimated by conducting a binary classification in $\{+1, -1\}$ with input $\boldsymbol{\tilde{y}}^f+2k\cdot \boldsymbol{\tilde{h}}_n^f$, which is denoted as
\begin{gather}
\label{eq:adnn_multi_class_ratio}
{{P_n\{{\hat{x}}_n^f = -2k+1| \boldsymbol{\tilde{y}}^f\}}\over
{P_n\{{\hat{x}}_n^f = -2k-1 | \boldsymbol{\tilde{y}}^f\}}} \approx
{\mathcal L}_b^{+-}(\boldsymbol{\tilde{y}}^f+2k\cdot \boldsymbol{\tilde{h}}_n^f).
\end{gather}
In this way, the multi-class classification is transformed into multiple binary detection processes.
For ease of discussion, the operation $\boldsymbol{\tilde{y}}^f+2k\cdot \boldsymbol{\tilde{h}}_n^f$ is referred to as the ``shifting process" for the remainder of this paper, where the step size $2k$ is referred to as the ``shifting parameter" and denoted as $s_n$.
Note that while the above transformation is designed for QAM modulation, the same idea can be generalized to other modulation schemes with a customized shifting process.

At the testing time, the received signal is tested along with all the possible values of the shifting parameter $s_n$ in set $\mathcal{S} = \{-2K, -2K+2, \dots, 2K-2, 2K\}$.
The probability of each class can be obtained by collecting all the likelihood ratios in eq.(~\ref{eq:adnn_multi_class_ratio}) with
\begin{multline}
    P_n\{{\hat{x}}_n^f = -2k+1| \boldsymbol{\tilde{y}}^f\}\\
    \approx P_n\{{\hat{x}}_n^f \!=\! -2K-1| \boldsymbol{\tilde{y}}^f \} \!\prod_{k'=k}^{K}\!{\mathcal L}_b^{+-}(\boldsymbol{\tilde{y}}^f + 2k'\cdot \boldsymbol{\tilde{h}}_n^f),
\end{multline}
where each probability can be computed by solving the equation $\sum_{a \in \mathcal{A}} P_n\{{\hat{x}}_n^f = a| \boldsymbol{\tilde{y}}^f\} = 1$.
The estimated ${\hat{x}}_n^f$ is determined by the class with the maximum probability.
In Fig.~\ref{figs:adnn_structure}, we show the architecture of ADNN and an example of when it is tested in the $16$-QAM case.
The ADNN is employed as the frequency domain network in RC-Struct, where the channel $\boldsymbol{\tilde{h}}_n^f$ is estimated by the linear minimum mean square error (LMMSE) method.

\subsubsection{Construction of binary training samples}
\label{sec:structnet_analysis_train_symbol_construction}

The complex-valued $M$-QAM symbols are converted to real-valued $\sqrt{M}$-PAM symbols for training. 
For clarity, we define two types of training samples: the PAM training samples and the binary training samples.
The PAM training samples refer to the real-valued transmit and receive signal pairs.
The binary training samples are generated to train the designed binary classifier using the PAM training samples.
For each PAM training sample, we construct two binary training samples, which include a positive binary sample and a negative binary sample.
As mentioned in Sec.~\ref{sec:prelim_adnn_multi_class}, the binary label is determined by $b_n = s_n + {\tilde{x}}_n^f$. 
At the training time, the shifting parameter is set as $s_n = -{\tilde{x}}_n^f+ b_n$ to control if a binary sample is positive or negative.
Specifically, for each PAM training sample $({\tilde{x}}_n^{f}, \boldsymbol{\tilde{y}}^f)$, positive and negative binary training samples are generated as 
\begin{align}
    (b_n^{+}=+1, \boldsymbol{\tilde{y}}^{f}, s_n^{+} =-{\tilde{x}}_n^{f}+1), \nonumber\\
    (b_n^{-}=-1, \boldsymbol{\tilde{y}}^{f}, s_n^{-} =-{\tilde{x}}_n^{f}-1).
\end{align}
In this way, each PAM training sample is augmented into two binary training samples and thus can be utilized more efficiently.
In Tab.~\ref{tab:data_construct_example}, we show examples of positive and negative shifting parameters corresponding to $4$-PAM symbols.


\begin{table}
\centering
\caption{Construction of binary training sample}
\begin{tabular}{ccc}
\toprule
Transmit symbol $\tilde{x}_n^f$ & $s_n^{+} = -\tilde{x}_n^f+1$ & $s_n^{-} = -\tilde{x}_n^f-1$ \\
\midrule
$-3$ & $4$ & $2$ \\
$-1$ & $2$ & $0$ \\
$+1$ & $0$ & $-2$ \\
$+3$ & $-2$ & $-4$ \\

\bottomrule
\end{tabular}
\vspace{-1em}
\label{tab:data_construct_example}
\end{table}

\section{Frequency domain network --- StructNet}
\label{sec:structnet_analysis}

In this section, we introduce the frequency domain network, StructNet, in a MIMO system and analyze the properties of the network.
The discussion of how to apply StructNet in the MIMO-OFDM symbol detection task is provided in Sec.~\ref{sec:introduced_method}.

\subsection{Design of StructNet}
\label{sec:structnet_analysis_design}
StructNet is a frequency domain network that builds upon ADNN, where it also embeds the repetitive structure of the modulation constellation into the network.
In ADNN, the shifting process is performed using the ground truth channel.
When adopted in RC-Struct, the LMMSE estimated channel is utilized due to the difficulty in obtaining perfect channel knowledge in practice. 
While the estimated channel can serve as an alternative to conduct the shifting process, it lacks the ability to adapt to changes in the environment. To address this issue, an additional PE layer is introduced in StructNet to estimate channel and perform the shifting process. The weights of the PE layer are dynamically updated according to channel variations without relying on perfect channel knowledge.

\begin{figure}
\centering
\includegraphics[width=0.8\linewidth]{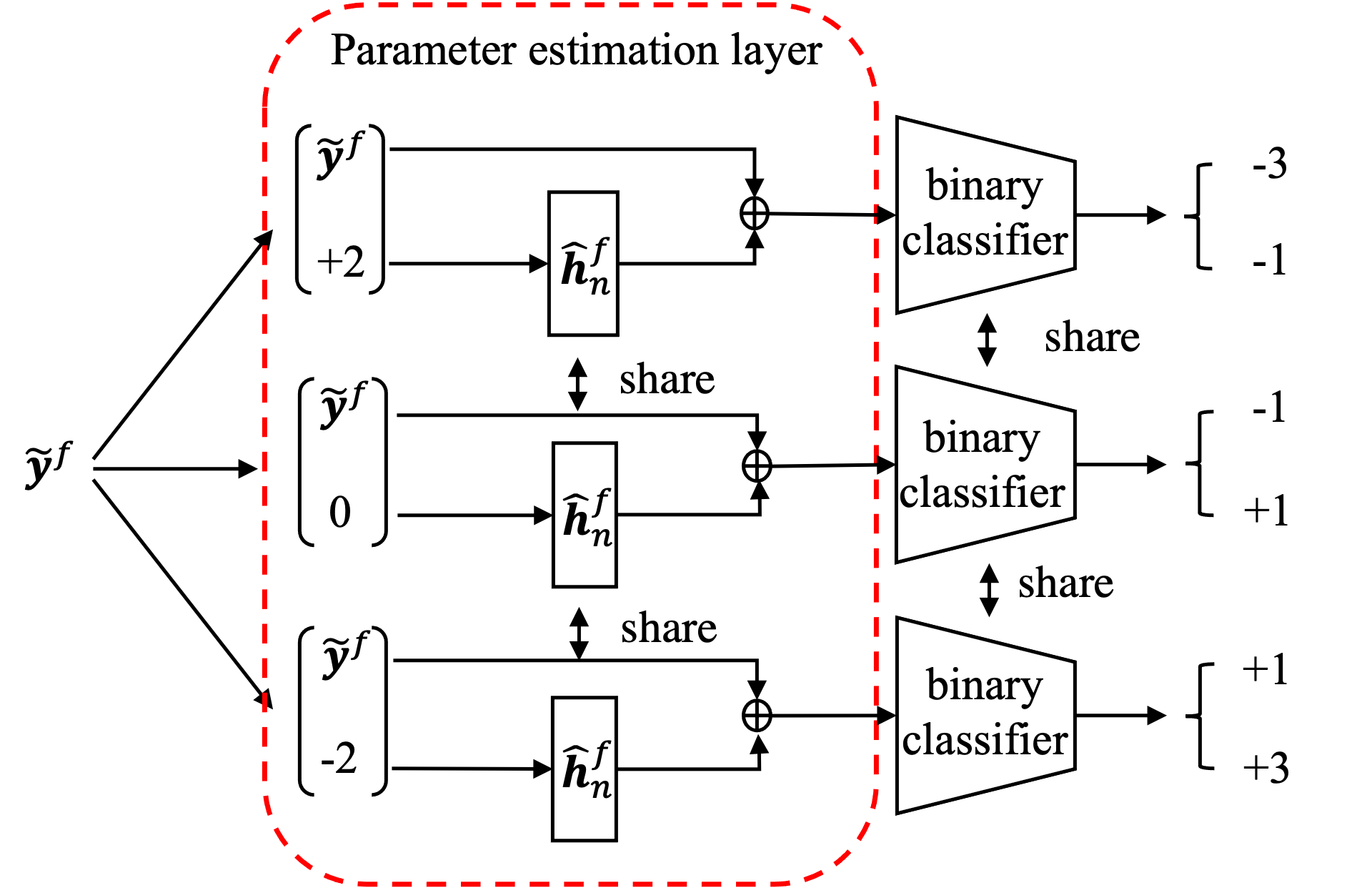}
\caption{The architecture of StructNet.}
\label{figs:structnet_test}
\vspace{-1em}
\end{figure}

The architecture of StructNet is depicted in Fig.~\ref{figs:structnet_test}, comprising a PE layer and a binary classifier. 
The weights of the PE layer, denoted by $\boldsymbol{\hat{h}}_n^f$, also represent an estimate of the ground truth channel $\boldsymbol{\tilde{h}}_n^f$. 
The PE layer takes the shifting parameter $s_n \in \mathcal{S}$ as input and is implemented using a linear NN layer. 
The binary classifier is an MLP. 
The PE layer is initialized with the LMMSE estimated channel and updated along with the binary classifier through backpropagation.
The network is trained using the cross-entropy loss.

One notable aspect of StructNet is that its PE layer for channel estimation is learned without assuming knowledge of the ground truth channel. 
Unlike other NN-based channel estimation approaches, which require ground truth channels to calculate the network loss, StructNet adopts a more realistic setting and does not assume perfect CSI. 
Instead, the PE layer is trained based on the loss of the binary classification. 
Such an update of the PE layer is achieved by embedding the repetitive modulation constellation structure into the network design.
In addition, the embedded structural information also enables StructNet to utilize training samples more efficiently, as discussed in Sec.\ref{sec:structnet_analysis_train_symbol_construction}, and to be more robust to incorrect labels, as analyzed in Sec.\ref{sec:structnet_analysis_incorrect_labels}. 
The robustness to incorrect labels makes StructNet less susceptible to error propagation when applied in the DF mechanism. 
Furthermore, the introduced PE layer further facilitates the DF approach due to the dynamic updating of the network parameters.

\begin{figure*}
\centering
\includegraphics[width=0.5\linewidth]{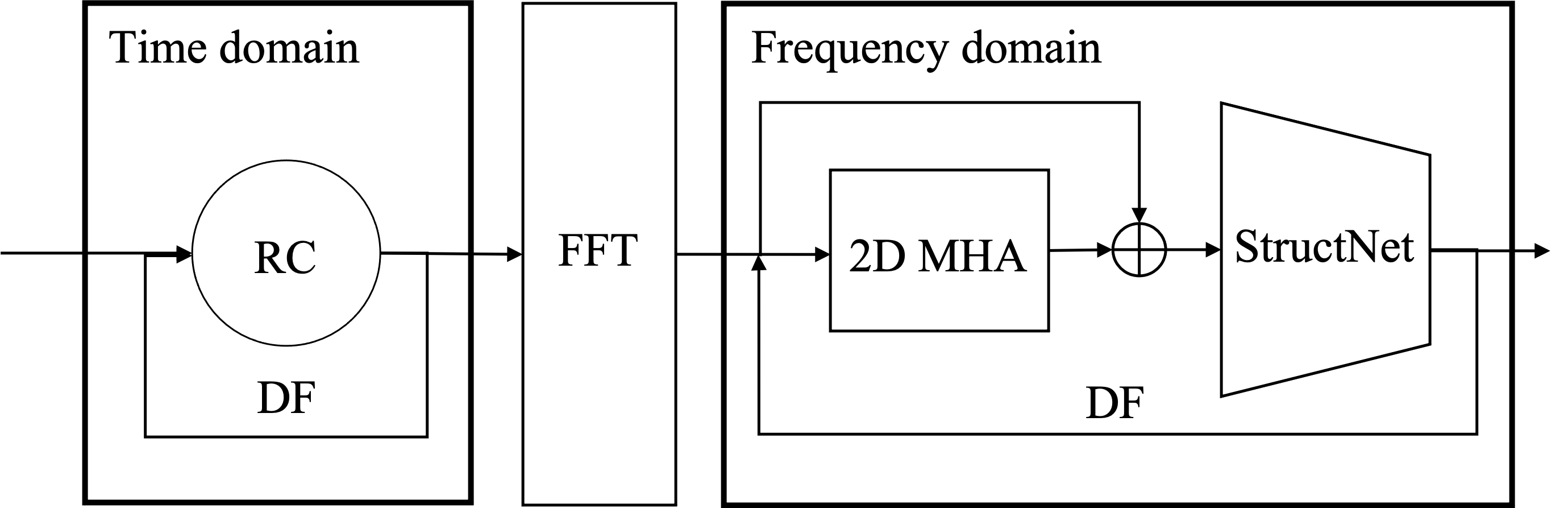}
\caption{Architecture of RC-AttStructNet-DF network.}
\label{figs:network_architecture}
\vspace{-1em}
\end{figure*}

\subsection{Training with incorrect labels}
\label{sec:structnet_analysis_incorrect_labels}
Due to the unique training sample construction process discussed in Sec.~\ref{sec:structnet_analysis_train_symbol_construction}, given any percentage of the incorrect PAM labels, there are at most $50\%$ incorrect binary labels during the training stage.
We first explain the reason for this conclusion with a concrete example and then discuss it in a more general case.



For ease of discussion, we use $4$-PAM as an example, but the conclusion generalizes to $\sqrt{M}$-PAM labels.
Note that there are two kinds of labels in our setting: the PAM labels and the binary labels.
The PAM label is the transmitted $\sqrt{M}$-PAM symbol in the set $\mathcal{A}$.
The binary label is in the set $\{+1, -1\}$, indicating whether a sample is positive or negative.
Assume we have a correct PAM label with ${\tilde{x}}_n^f = -1$ and the corresponding incorrect PAM label is ${\bar{x}}_n^f = +1$.
Then the training samples created by the incorrect label ${\bar{x}}_n^f = +1$ can be written as
\begin{align}
    (\bar{b}_n^{+}=+1, \boldsymbol{\tilde{y}}^f, {s}_n^{+} =-{\bar{x}}_n^f+1=0), \nonumber\\
    (\bar{b}_n^{-}=-1, \boldsymbol{\tilde{y}}^f, {s}_n^{-} =-{\bar{x}}_n^f-1=-2).
\end{align}
Note that the actual binary label is determined by ${b}_n = s_n + {\tilde{x}}_n^f$.
As the actual PAM label is ${\tilde{x}}_n^f = -1$, the actual binary label for the constructed positive sample, however, is ${b}_n^{+} = s_n^{+} + {\tilde{x}}_n^f = -1 < 0$, resulting in an incorrect binary positive sample.
For the negative sample, the actual binary label is ${b}_n^{-} = s_n^{-} + {\tilde{x}}_n^f = -3 < 0$, which is a correct binary sample.

In general, suppose ${\bar{x}}_n^f$ is the incorrect PAM label corresponding to the actual PAM label ${\tilde{x}}_n^f$.
The positive sample is created by setting $s_n^{+} = -{\bar{x}}_n^f + 1$ and the negative sample is constructed by setting $s_n^{-} = -{\bar{x}}_n^f - 1$.
As the actual PAM label is ${\tilde{x}}_n^f$, the actual binary label for the positive sample is ${b}_n^{+} = {\tilde{x}}_n^f  + s_n^{+} = {\tilde{x}}_n^f - {\bar{x}}_n^f + 1$ and the actual label for the negative sample is ${b}_n^{-} = {\tilde{x}}_n^f  + s_n^{-} = {\tilde{x}}_n^f - {\bar{x}}_n^f - 1$.
The positive sample is considered to be incorrect if ${b}_n^{+} = {\tilde{x}}_n^f - {\bar{x}}_n^f + 1 < 0$, i.e., ${\tilde{x}}_n^f - {\bar{x}}_n^f < -1$.
The negative sample is incorrect when ${b}_n^{-} = {\tilde{x}}_n^f - {\bar{x}}_n^f - 1 > 0$, i.e., ${\tilde{x}}_n^f - {\bar{x}}_n^f > 1$.
Since ${\tilde{x}}_n^f, {\bar{x}}_n^f \in \mathcal{A}$ and ${\tilde{x}}_n^f \neq {\bar{x}}_n^f$, the distance ${\tilde{x}}_n^f - {\bar{x}}_n^f$ can at most satisfy one of the inequalities between ${\tilde{x}}_n^f - {\bar{x}}_n^f < -1$ and ${\tilde{x}}_n^f - {\bar{x}}_n^f > 1$ for any value of ${\tilde{x}}_n^f$ and ${\hat{x}}_n^f$.
Thus, there is at least one correct positive sample or correct negative sample for any incorrect PAM sample.
The percentage of incorrect binary samples is at most $50\%$.
The incorrect percentage only equals $50\%$ when all the training PAM labels are incorrect.
This unique property of StructNet makes it less affected by detection errors.
As one of the major issues of DF-based approaches is error propagation, StructNet can mitigate the issue by showing robustness to the detection errors and therefore work well with the DF approach.

\section{Introduced approach --- RC-AttStructNet-DF}
\label{sec:introduced_method}

In this section, we introduce the RC-AttStructNet-DF approach. 
As shown in Fig.~\ref{figs:network_architecture}, the architecture of RC-AttStructNet-DF consists of three essential components: the time domain RC network, the 2D MHA module, and the frequency domain StructNet network.
The time domain RC network is exploited to decouple the transmitted data streams and deconvolve the channel for equalization~\cite{zhou2019, zhou2020rcnet}.
Furthermore, we design a DF mechanism to dynamically update the network with the detected data symbols, especially in high mobility scenarios. 
The RLS algorithm is used in the time domain as part of the DF mechanism to adaptively update the weights on an OFDM symbol basis. 
In the frequency domain, the 2D MHA module is designed to capture the time and frequency correlation of the signal.
Subsequently, the frequency domain network, StructNet, is utilized to conduct the multi-class classification.
The residual connection~\cite{he2016deep} is employed to connect the 2D MHA and the StructNet.
Additionally, we develop an attention loss for the DF mechanism in the frequency domain, which weights the training loss of different samples based on their confidence level. 
In the following subsections, we provide detailed training and testing procedures for our approach.




\subsection{RC with DF}
\label{sec:rc_with_gaw_rls}
In the time domain, the convolution and the superposition operation of the wireless channel are conducted on the transmit signal. 
The RC reverses such an operation by jointly decoupling the different transmitted data streams and deconvolving the channel for equalization~\cite{zhou2019, zhou2020rcnet}.

\subsubsection{Learning from pilot symbols}
The input to the RC network is the time domain received signal $\boldsymbol{Y}^t$ and the target output is the time domain transmitted signal $\boldsymbol{X}^t$.
The training dataset for learning from pilot symbols can be represented as
\begin{align}
\label{eq:rc_pilot_train_data}
    \mathcal{D}_{rc} \triangleq \{ \boldsymbol{Y}_n^t, \boldsymbol{X}_n^t \}_{n = 0}^{N_p - 1}.
\end{align}
The initial output weights of RC are learned with the pilot symbols through the LS solution shown in eq.~(\ref{equ:rc_least_square}).

\subsubsection{DF with data symbols}
The DF mechanism is utilized to dynamically update the output weights of RC with the detected data symbols in a symbol-by-symbol manner.
During the DF procedure, the RLS~\cite{farhang2013adaptive, jaeger2002adaptive} method is adopted to recursively update the output weights.
The initial weights for the RLS algorithm are the weights learned from the pilot symbols.

Specifically, we use RC trained by the pilot symbols to generate the estimation for the first data symbol $\boldsymbol{\hat{X}}_{N_p}^t \in {\mathbb C}^{N_t \times (N_{\mathrm{cp}}+N_{\mathrm{sc}})}$.
For the $n$th data symbol, the estimation $\boldsymbol{\hat{X}}_n^t$ ($n = N_p + 1, \dots, N - 1$) is detected by the RC learned with the $(n-1)$-th data symbol.
The estimation is then converted into frequency domain $\boldsymbol{\hat{X}}_n^f$ and then each symbol is mapped to the nearest constellation points, which generates $\boldsymbol{\bar{X}}_n^f$.
The frequency domain $\boldsymbol{\bar{X}}_n^f$ is transformed back to the time domain $\boldsymbol{\bar{X}}_n^t$ through the IFFT operation.
Then the output weights of RC are recursively updated by minimizing the objective:
\begin{align}
\label{eq:rls_weight_update}
    \arg\min_{\boldsymbol{W}_{\mathrm{out}}^{(n)}(m)} \sum_{m'=0}^{m} \alpha^{m - m'} \| \boldsymbol{W}_{\mathrm{out}}^{(n)}(m')\boldsymbol{z}_n(m') - \boldsymbol{\bar{x}}_n^t(m') \|_2^2,
\end{align}
where $\boldsymbol{W}_{\mathrm{out}}^{(n)}(m)$ is the weight learned by the $n$-th data symbol at step $m$; $\boldsymbol{z}_n(m) \in \mathbb{C}^{N_n+N_t}$ is the concatenation of the RC state and input for the $n$-th data symbol at step $m$; $\boldsymbol{\bar{x}}_n^t(m) \in {\mathbb C}^{N_t}$ is the $m$-th column of $\boldsymbol{\bar{X}}_n^t$; $\alpha \in (0, 1]$ is the forgetting factor; and $m \!=\! 0, 1, \!\dots\!, N_{\mathrm{sc}}\! +\! N_{\mathrm{cp}} \!-\! 1$.
The forgetting factor $\alpha$ indicates how much we trust the previous samples.
When $\alpha < 1$, the smaller the $\alpha$, the fewer weights we put on the old samples than the recent ones.

The output weight $\boldsymbol{\hat{W}}_{\mathrm{out}}^{(n)}(m)$ is recursively updated by
\begin{align}
\label{equ:gaw_rls_wout_update}
    \boldsymbol{\hat{W}}_{\mathrm{out}}^{(n)}(m) = \boldsymbol{\hat{W}}_{\mathrm{out}}^{(n)}(m-1) + \boldsymbol{e}_{n}(m)\boldsymbol{v}_n^T(m),
\end{align}
where $\boldsymbol{e}_{n}(m) = \boldsymbol{\bar{x}}_n^t(m) - \boldsymbol{\hat{W}}_{\mathrm{out}}^{(n)}(m-1)\boldsymbol{z}_n(m)$ is the error on the current sample $m$ when estimated with weight matrix at the $(m-1)$-th step, and $\boldsymbol{v}_n(m)$ is the gain vector computed by the following equation~\cite{jaeger2002adaptive}:
\begin{align}
\label{equ:gaw_rls_k_m_update}
    \boldsymbol{v}_n(m) = \frac{\boldsymbol{\Phi}_n^{-1}(m-1)\boldsymbol{z}_n(m)}
    {\alpha + \boldsymbol{z}_n^T(m)\boldsymbol{\Phi}_n^{-1}(m-1)\boldsymbol{z}_n(m)}.
\end{align}
The $\boldsymbol{\Phi}_n^{-1}(m) = (\sum_{m'=0}^{m} \alpha^{m - m'}\boldsymbol{z}_n(m')\boldsymbol{z}_n^T(m'))^{-1}$ is the inverse of the weighted correlation of $\boldsymbol{z}_n(m)$ and is recursively updated with
\begin{align}
\label{equ:gaw_rls_phi_update}
    \boldsymbol{\Phi}_n^{-1}\!(m) \!=\! \alpha^{-1}(\boldsymbol{\Phi}_n^{-1}(m-1) - \boldsymbol{v}_n(m)\left[\boldsymbol{z}_n^T(m)\boldsymbol{\Phi}_n^{-1}(m-1)\right]).
\end{align}
The output weight $\boldsymbol{\hat{W}}_{\mathrm{out}}^{(n)}$ is determined by the weight learned at $(N_{\mathrm{sc}}\! +\! N_{\mathrm{cp}} \!-\! 1)$-th step.


\subsection{2D MHA and StructNet with DF}

In the frequency domain, the output of the RC $\boldsymbol{\hat{X}}_n^t$ is transformed to the frequency domain $\boldsymbol{\hat{X}}_n^f \in \mathbb{C}^{N_t \times N_{\mathrm{sc}}}$, where the corresponding target output is $\boldsymbol{X}_n^f$.
Denote ${\hat{X}}_n^f(n_t, n_{\mathrm{sc}})$ and ${X}_n^f(n_t, n_{\mathrm{sc}})$ as the $(n_t, n_{\mathrm{sc}})$-th entry of the $\boldsymbol{\hat{X}}_n^f$ and $\boldsymbol{X}_n^f$.
The complex values are mapped to real values by $\boldsymbol{i}_n^f(n_t,n_{\mathrm{sc}})=f_R({\hat{X}}_n(n_t, n_{\mathrm{sc}}))$ and $\boldsymbol{o}_n^f(n_t,n_{\mathrm{sc}}) = f_R({{X}}_n(n_t, n_{\mathrm{sc}}))$.

\subsubsection{Learning from pilot symbols}

\begin{figure}
\centering
\includegraphics[width=0.6\linewidth]{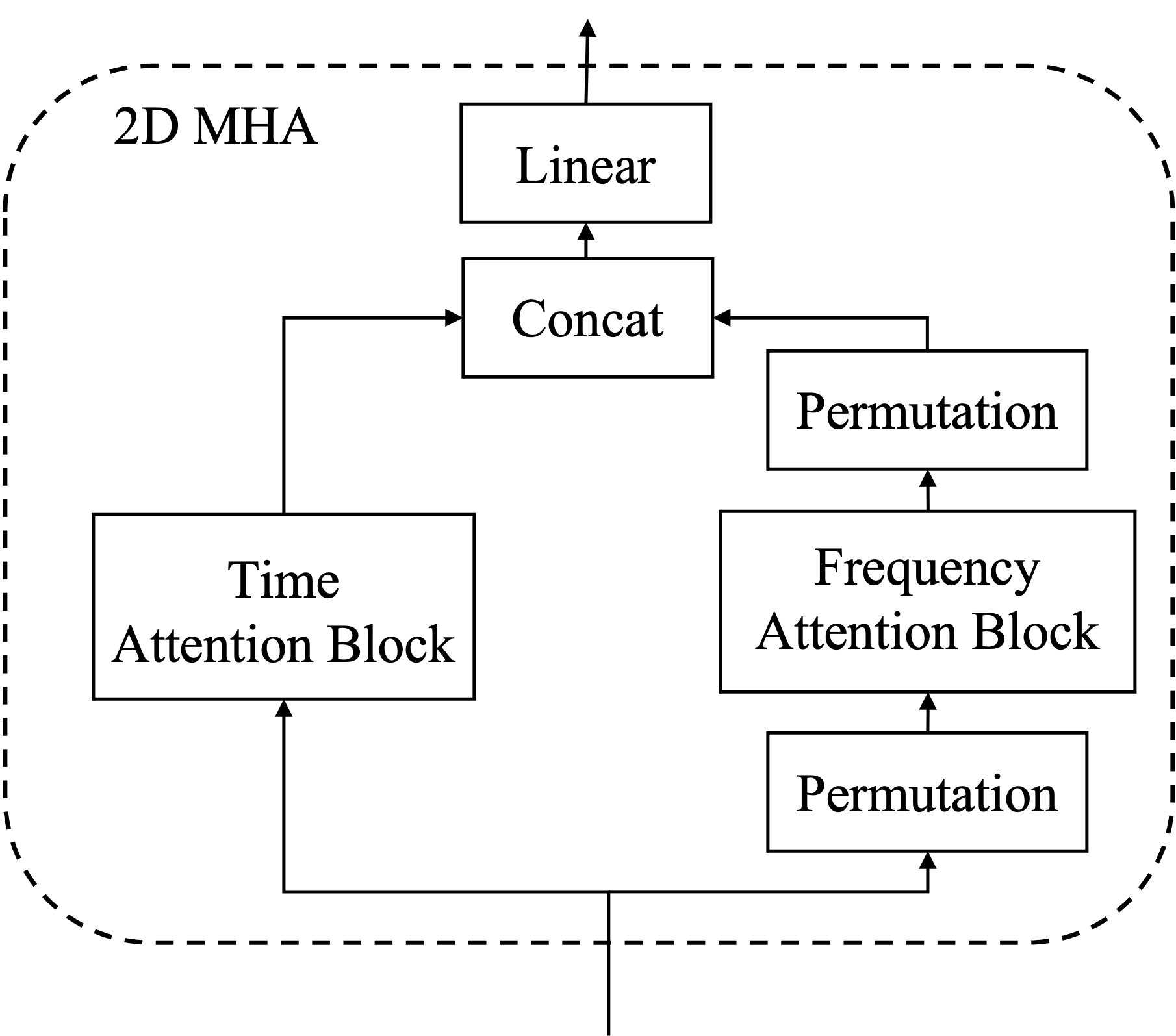}
\caption{2D MHA module.}
\label{figs:2d_mha_architecture}
\vspace{-1em}
\end{figure}

The input first passes through the 2D MHA module.
Unlike the MHA module in the Transformer that only captures feature correlations along the same dimension, the 2D MHA works in a two-dimensional manner, incorporating both time and frequency dimensions.
As shown in Fig.~\ref{figs:2d_mha_architecture}, two attention blocks, including a time attention block and a frequency attention block, are utilized in the module to capture the time correlation and frequency correlation, respectively.
The input to the time attention block is the real-valued frequency domain signal $\boldsymbol{I}^f(n_t) \in \mathbb{R}^{N_p\times 2N_{\mathrm{sc}}}$, which is a concatenation of $\boldsymbol{i}_n^f(n_t,n_{\mathrm{sc}})$ along the time and frequency dimensions.
The input to the frequency attention block is 
the frequency domain signal $\boldsymbol{\tilde{I}}^f(n_t) \in \mathbb{R}^{N_{\mathrm{sc}}\times 2N_{p}}$ obtained by a permutation of $\boldsymbol{I}^f(n_t)$.
To facilitate the feature aggregation of each block, the value embedding dimension for each attention block is set to be equal to the input size.
The output of the frequency attention block is permutated and then aggregated with the time attention block output through concatenation and linear projection.
The final output of 2D MHA is added to the input through the residual connection and then passed through StructNet.
Note that the 2D MHA is only employed when learning from pilot symbols to help obtain a better estimate of the data symbols.
During the DF procedure, the network is updated on a symbol basis, which does not allow the module to capture the two-dimensional feature along the time dimension.
Therefore, the 2D MHA is skipped during the DF procedure.

\subsubsection{Attention loss}
The attention loss is designed for the frequency domain network to re-weight the training loss of different samples according to their confidence level.
The idea behind it is to force the network to pay more attention to confident samples.
The attention loss for each sample can be written as $\boldsymbol{a}_n^f(n_t,n_{\mathrm{sc}})\odot\ell(f_s(\boldsymbol{i}_n^f(n_t,n_{\mathrm{sc}})), \boldsymbol{{o}}_n^f(n_t,n_{\mathrm{sc}}))$, where $\boldsymbol{a}_n^f(n_t,n_{\mathrm{sc}})$ is the corresponding weights of this training sample, $\odot$ is the Hadamard product, $\ell(\cdot)$ stands for the cross-entropy loss, and $f_s(\cdot)$ represents the function approximated by the frequency domain network.
When training with the pilot symbols, the confidence level for all the training samples is the same, and thus the weights for all the training samples are set to be 1, i.e., $\boldsymbol{a}_n^f(n_t,n_{\mathrm{sc}}) = [1, 1]^T$.
Note that the output of the frequency domain network is the detected data symbol along with its predicted probability.
The predicted probability indicates how confident the network is in the detected data symbol.
During the DF procedure, the training labels are the detected symbols.
The attention weights are set as the predicted probability of detected symbols.
It is true that the detected data symbol with a high predicted probability is not ensured to be correct. 
However, if the network predicts a high probability for the detected data symbol, it is more likely to be correct than the detected data symbol with a low predicted probability. 
When the detected data symbol has a high predicted probability, we put more weights on it than the detected symbol with a low predicted probability.
In addition, only symbols with a predicted probability larger than $\eta$ are used during the DF process. Therefore, the attention weights can be written as $\boldsymbol{a}_n^f(n_t,n_{\mathrm{sc}}) = q(P(\boldsymbol{\tilde{o}}_n^f(n_t,n_{\mathrm{sc}})))$, where $P(\boldsymbol{\tilde{o}}_n^f(n_t,n_{\mathrm{sc}}))$ is the probability of the predicted symbol $\boldsymbol{\tilde{o}}_n^f(n_t,n_{\mathrm{sc}})$ provided by the frequency domain network. 
The $q(\cdot)$ denotes the following non-linear function:
\begin{align}
    q(x)= 
\begin{cases}
    x,& \text{if } x\geq \eta \\
    0,              & \text{otherwise}
\end{cases},
\end{align}
where $\eta$ is the probability threshold.
With the attention loss, the error propagation issue of the DF process can be alleviated by assigning weights to the loss based on the predicted probability and selecting samples with high confidence.

\subsubsection{DF with data symbols}
After training the 2D MHA and the StructNet with the pilot symbols, we adopt the DF mechanism to learn from the detected data symbols on an OFDM symbol basis.
As the two-dimensional features do not exist when updating in a symbol-by-symbol fashion, the 2D MHA module is not used and only the StructNet is fine-tuned with the data symbols.
Specifically, we obtain the detected data signals $\boldsymbol{\hat{X}}_n^t$ in the time domain utilizing the updated RC and transform it to frequency domain $\boldsymbol{\hat{X}}_n^f$. 
Then we test the StructNet updated by the $n-1$th data symbol with $\boldsymbol{\hat{X}}_n^f$ to get the estimated symbol $\boldsymbol{\tilde{X}}_n^f$ for the $n$th data symbol.
The pair $(\boldsymbol{\hat{X}}_n^f, \boldsymbol{\tilde{X}}_n^f)$ is exploited as the training data to fine-tune the StructNet.
Note that for $n=N_p$, i.e., the first data symbol, the training label is obtained by testing with the pilot-trained 2D MHA and the StructNet.
For the rest data symbols, the training labels for fine-tuning are obtained only by the fine-tuned StructNet.

\subsection{Summary of symbol detection procedure}

The symbol detection procedure of RC-AttStructNet-DF includes two parts: offline initialization and online training.
We explain the offline initialization and summarize the online training in this subsection.
The procedure is also shown in \textbf{Algorithm~\ref{alg:rc_structnet_df}}.

\begin{algorithm}
\caption{Symbol Detection Procedure of RC-AttStructNet-DF}\label{alg:rc_structnet_df}
\footnotesize
\begin{algorithmic}[1]
    \commentline{Offline initialization}
    \State Prepare artificial training data
    \State Train a binary classifier with artificial training data
    \State Initialize binary classifier in StructNet with learned weights
    \commentline{Online training}
    \For{Each OFDM subframe}
        \commentline{Learning from pilot symbols}
        \State \multiline{Prepare pilot training data $D_{rc}$ with time domain received and transmitted pilot signals as defined in eq.~(\ref{eq:rc_pilot_train_data})}
        \State \multiline{Train output weights of RC with $D_{rc}$ using one-shot LS solution in eq.~(\ref{equ:rc_least_square})}
        \State \multiline{Generate RC output (time domain RC pilot output) when tested on received pilot signals}
        \State Transform time domain RC pilot output to frequency domain (frequency domain RC pilot output) 
        \State Utilize frequency domain RC pilot output and transmitted pilot symbols to estimate effective channel using LMMSE
        \State Initialize PE layer of StructNet with estimated effective channel
        \State Prepare pilot training data for 2D MHA and StructNet with frequency domain RC pilot output and transmitted pilot symbols
        \State Train 2D MHA and StuctNet with pilot training data using attention loss
        \commentline{DF with data symbols}
        \For{OFDM symbol $n = N_p: N-1$ (data symbols)}
            \State Generate RC output ($n$-th time domain RC data output) by testing on $n$-th time domain received data signal
            \State Prepare training dataset for RC with $n$-th time domain received data signal and $n$-th time domain RC data output
            \State Update RC weights by RLS algorithm following eq.~(\ref{eq:rls_weight_update})
            \State Transform $n$-th time domain RC data output to frequency domain ($n$-th frequency domain RC data output)
            \If{$n=N_p$}
                \State \multiline{Generate $n$-th detected data symbols with 2D MHA and StructNet by testing on $n$-th frequency domain RC data output}
            \Else
                \State \multiline{Generate $n$-th detected data symbols with StructNet by testing on $n$-th frequency domain RC data output}
            \EndIf
            \State \multiline{Prepare training dataset for StructNet with $n$-th frequency domain RC data output and $n$-th detected data symbols }    
            \State Fine-tune StructNet weights with attention loss
        \EndFor
    \EndFor
\end{algorithmic}
\end{algorithm}

\subsubsection{Offline initialization}
The binary classifier in StructNet is initialized with offline weights that are trained by artificially generated data.
Note that this offline training does not require any prior knowledge of the channel, which differentiates our method from other work with offline initialization and online adaptation.
Specifically, the training labels are randomly generated symbols $\boldsymbol{E} \in \mathcal{A}^{N_t \times N_{sp}}$, where $\mathcal{A}$ represents the set $\{-1,+1\}$ and $N_{sp}$ is the number of offline training samples.
The training inputs are the noise-contaminated symbols or the received signal $\boldsymbol{E}+\boldsymbol{G}$, where $\boldsymbol{G} \in \mathbb{R}^{N_t \times N_{sp}}$ is the Gaussian noise.
In this way, the initial weights of the binary classifier are learned to conduct nearest neighbor mapping and facilitate the online training of the full StructNet, i.e., the PE layer and the binary classifier. 
Note that these offline weights, once trained, are fixed. 
The same weights are used for initialization when conducting online detection on different subframes.

\begin{table*}
\centering
\caption{Training complexity}
\resizebox{\linewidth}{!}{
\begin{tabular}{lcc}
\toprule
Method & RC & Frequency domain network \\
\midrule
RC-Struct & $\mathcal{O}(V(N_n + N_{\mathrm{train}} + N_t)(N_n+N_r)N_{\mathrm{train}})$ & $\mathcal{O}(8N_tN_{h}N_{\mathrm{ep}}N_{\mathrm{sc}}N_p)$ \\
RC-AttStructNet-DF & $\mathcal{O}(V(N_n + N_{\mathrm{train}} + N_t)(N_n+N_r)N_{\mathrm{train}} + ((N_n+N_r)N_t + 4(N_n+N_r)^2)N_{\mathrm{test}})$ & $\mathcal{O}(4N_pN_{\mathrm{sc}}N_t(2N_k+N_{\mathrm{sc}}) + 8N_t(N_{\mathrm{ep}}N_p + N_{\mathrm{epdf}}N_d)N_{h}N_{\mathrm{sc}})$ \\
\bottomrule
\end{tabular}
}
\label{tab:train_complex_table}
\end{table*}

\begin{table*}
\centering
\caption{Testing complexity}
\begin{tabular}{lc}
\toprule
Method & Complexity \\
\midrule
LMMSE with LMMSE-CSI& $\mathcal{O}(N_pN_{\mathrm{sc}}^2N_a^2 + N_dN_{\mathrm{sc}}(N_a^3+N_a^2+N_a))$ \\
LMMSE with LMMSE-Interpolation & $\mathcal{O}((N_pN_{\mathrm{sc}}^2 + 7N_dN_{\mathrm{sc}})N_a^2 + N_dN_{\mathrm{sc}}(N_a^3+N_a^2+N_a))$ \\
SD with LMMSE-Interpolation & $\mathcal{O}((N_pN_{\mathrm{sc}}^2 + 7N_dN_{\mathrm{sc}})N_a^2 + N_dN_{\mathrm{sc}}M^{N_a}(2N_a^2+2N_a-1))$ \\
RC-Struct & $\mathcal{O}(V(N_n+N_r)(N_nN_{\mathrm{test}} + N_t)+4N_tN_{h}N_{\mathrm{sc}}N_d)$ \\
RC-AttStructNet-DF & $\mathcal{O}(V(N_n+N_r)(N_nN_{\mathrm{test}} + N_t)+4N_{\mathrm{sc}}N_t(2N_k+N_{\mathrm{sc}} + N_{h}N_d))$ \\

\bottomrule
\end{tabular}
\label{tab:test_complex_table}
\end{table*}

\subsubsection{Online training}
The time domain RC network and the frequency domain network are learned separately. 
When learning from pilot symbols, RC is first trained with the time domain received and transmitted pilot signals.
Then after the training of RC, the output of RC is transformed into the frequency domain.
The frequency domain network is learned by taking frequency domain RC output as the input and the transmitted pilot symbols in the frequency domain as the training label.
When learning from detected data symbols with the DF mechanism, the network weights are updated in a symbol-by-symbol manner.
For each data symbol, we generate the RC output of the data symbol by testing on the time domain received data signal.
This RC output is employed as the training label to re-train RC with the RLS algorithm.
The time domain RC output is then converted to frequency domain.
The detected data symbol is obtained by testing the frequency domain network on frequency domain RC output.
Similarly, this detected data symbol is utilized as the training label for fine-tuning frequency domain network.


It is noteworthy that the combination of RC in the time domain and the frequency domain network is critical in our design. 
This is because even though StructNet is designed to learn from the training samples efficiently, training StructNet along with 2D MHA still requires a relatively large amount of training data. 
Since RC can efficiently decouple different data streams and equalize the channel in time domain, as shown in our previous work~\cite{zhou2019, zhou2020rcnet}, the classification task in frequency domain becomes much more accessible to tackle after the processing of RC. 

\section{Complexity Analysis}
\label{sec:complexity_analysis}

This section analyzes the computational complexity of RC-AttStructNet-DF. 
The complexity is compared with RC-Struct~\cite{xu2021rcstruct}, LMMSE detector, and sphere decoding (SD) approach.
In the analysis, we mainly consider the computation cost of matrix multiplication and pseudo-inverse, as the costs for matrix addition and element-wise operation are negligible compared to these main factors.
For ease of discussion, we denote the number of training and testing samples in the time domain as $N_{\mathrm{train}} = (N_{\mathrm{cp}} + N_{\mathrm{sc}})N_p$ and $N_{\mathrm{test}} = (N_{\mathrm{cp}} + N_{\mathrm{sc}})N_d$.
As the complexities of RC-Struct have been provided in~\cite{xu2021rcstruct}, we summarize the conclusions in Tab.~\ref{tab:train_complex_table} and Tab.~\ref{tab:test_complex_table} and mainly focus on the complexity analysis of RC-AttStructNet-DF.
$V$ denotes the number of cascaded RCs.

In the time domain, RC is first learned with the pilot symbols and then updated with the detected data symbols.
As the DF is utilized, the training complexity will be larger than RC-Struct due to the adoption of the RLS update procedure with extra training on the data symbols.
The complexity for training RC with pilot symbols is the same as RC-Struct, which is $\mathcal{O}(V(N_n + N_{\mathrm{train}} + N_t)(N_n+N_r)N_{\mathrm{train}})$.
The RLS procedure on the data symbols has three steps to update the output weights.
The complexity for updating $\boldsymbol{\hat{W}}_{\mathrm{out}}^{(n)}(m)$ in eq.~(\ref{equ:gaw_rls_wout_update}) for each sample is $\mathcal{O}((N_n+N_r)N_t)$.
The update of $\boldsymbol{v}_n(m)$ in eq.~(\ref{equ:gaw_rls_k_m_update}) has a complexity of $\mathcal{O}((N_n+N_r)^2 + N_n+N_r) \approx \mathcal{O}((N_n+N_r)^2)$. 
The complexity for updating $\boldsymbol{\Phi}^{-1}_n(m)$ in eq.~(\ref{equ:gaw_rls_phi_update}) is $\mathcal{O}(3(N_n+N_r)^2)$.
Then the complexity for all the samples with the RLS approach is $\mathcal{O}(((N_n+N_r)N_t + 4(N_n+N_r)^2)N_{\mathrm{test}})$.
Thus, the training complexity in time domain is $\mathcal{O}(V(N_n + N_{\mathrm{train}} + N_t)(N_n+N_r)N_{\mathrm{train}} + ((N_n+N_r)N_t + 4(N_n+N_r)^2)N_{\mathrm{test}})$.
At the testing stage, time domain RC in RC-AttStructNet-DF has the same complexity as RC-Struct, which is $\mathcal{O}(V(N_n+N_r)(N_nN_{\mathrm{test}} + N_t))$.

The frequency domain network is composed of the 2D MHA module and the StructNet.
For simplicity, we assume the time and frequency attention block in the 2D MHA adopts the same key embedding dimension $N_k$.
Then the complexity for the time attention block is $\mathcal{O}((4N_kN_pN_{\mathrm{sc}} + 4N_pN_{\mathrm{sc}}^2)N_t)$.
The complexity for frequency attention block is $\mathcal{O}((4N_kN_pN_{\mathrm{sc}} + 4N_pN_{\mathrm{sc}})N_t)$.
The output linear project has a complexity of $\mathcal{O}(8N_pN_{\mathrm{sc}}N_t)$.
Thus, the total complexity for training 2D MHA with pilot symbols is $\mathcal{O}(4N_pN_{\mathrm{sc}}N_t(2N_k+N_{\mathrm{sc}}+3)) \approx 
 \mathcal{O}(4N_pN_{\mathrm{sc}}N_t(2N_k+N_{\mathrm{sc}}))$.
During the DF procedure, 2D MHA is only adopted for testing the first data symbol, resulting in a complexity of $\mathcal{O}(4N_{\mathrm{sc}}N_t(2N_k+N_{\mathrm{sc}}))$.

For StructNet, it consists of a PE layer and a binary classifier.
The PE layer conducts an element-wise multiplication and thus the complexity is ignored here.
Denote $N_h$ as the number of neurons in the input layer and $N_{\mathrm{ep}}$ as the number of training epochs.
When training with pilot symbols, the training complexity of StructNet is the same as the frequency domain network of RC-Struct, which is $\mathcal{O}(8N_tN_{h}N_{\mathrm{ep}}N_{\mathrm{sc}}N_p)$.
When DF is adopted, the extra training complexity is $\mathcal{O}(8N_tN_{h}N_{\mathrm{epdf}}N_{\mathrm{sc}}N_d)$, where $N_{\mathrm{epdf}}$ is the number of fine-tuning epochs for DF.
Thus, the total complexity is $\mathcal{O}(8N_t(N_{\mathrm{ep}}N_p + N_{\mathrm{epdf}}N_d)N_{h}N_{\mathrm{sc}})$.
As the number of testing samples in the frequency domain is $N_tN_{\mathrm{sc}}N_d$, the testing complexity is $\mathcal{O}(4N_tN_{h}N_{\mathrm{sc}}N_d)$.

\begin{figure*}
\centering%
\vspace{-1em}
\subfloat[]{\label{a}\includegraphics[width=0.35\linewidth]{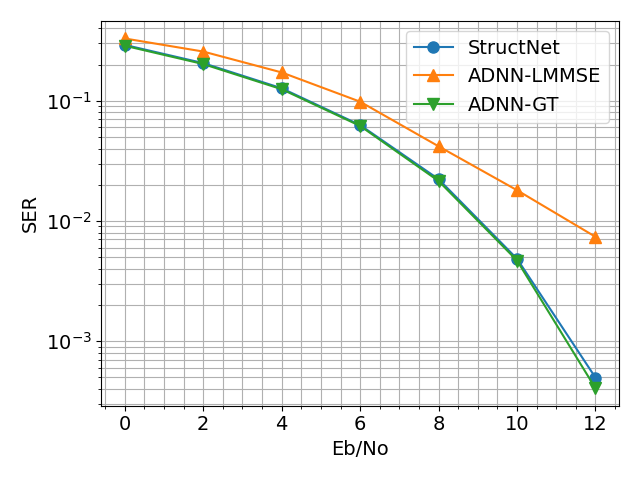}}
\hspace{3em}
\subfloat[]{\label{b}\includegraphics[width=0.35\linewidth]{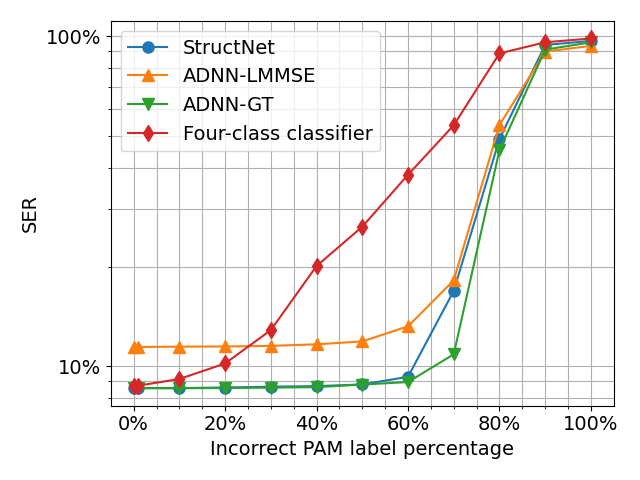}}
\caption{Comparison of SER.
(a) with different $Eb/No$'s in dB.
(b) with different percentages of incorrect PAM labels.
\label{figs:toy_example_ser}
}
\vspace{-1em}
\end{figure*}

The LMMSE approach is a low-complexity linear detector that is widely used in communication systems. 
The SD detector~\cite{ghasemmehdi2011faster} is a non-convex solver that approaches the optimal maximum likelihood (ML) detection, which has high detection complexity and thus is rarely used in practice.
As channel estimation is needed as the input to both methods, we adopt LMMSE for the underlying channel estimation. 
As our previous work~\cite{zhou2019} has analyzed the complexities of both methods with the LMMSE channel estimation in details, we summarize the conclusions in Tab.~\ref{tab:test_complex_table}.
The ``LMMSE-CSI" indicates that the channel estimates are obtained by only utilizing the pilot symbols. 
Meanwhile, the ``LMMSE-Interpolation" means that the channel estimates are interpolated over the data symbols using the pilot-estimated CSI.
To simplify the expression, we assume that the number of antennas satisfies $N_a = N_r = N_t$.

The analysis shows that RC-AttStructNet-DF has higher training and testing complexity than RC-Struct due to the additional 2D MHA module and DF procedure.
However, the training and testing complexities of these two approaches are still in the same order of magnitude.
Compared with the conventional approaches, RC-AttStructNet-DF has higher complexity than the LMMSE approach due to the extra training stage, and a lower complexity than the SD detector.

\section{Toy experiment: MIMO system with Gaussian channel}
\label{sec:toy_experiment_mimo}

In this section, we provide a toy experiment in a MIMO system with the Gaussian channel to analyze the properties of StructNet. 
We start with analyzing the effectiveness of the PE layer and then empirically show the robustness of StructNet to incorrect labels.

\subsection{Experimental setting}

In the toy experiment, we assume the classifier has sufficient data and time to be trained, and the PE layer is initialized with an inaccurate LMMSE estimated channel.
To satisfy such an assumption, the number of training samples is set to be $1000$, among which $4$ samples are utilized for LMMSE channel estimation, and the rest $996$ samples are used for training the classifier.
The trained network is tested with $3000$ samples.
For simplicity, we test in a $2\times2$ MIMO with $4$-PAM modulation.
$100$ Gaussian channel realizations are tested.
The channels are selected to have condition numbers smaller than $1.5$ to mimic the setting with dynamic transmission modes, where data is transmitted in channels with small condition numbers.
In Sec.~\ref{sec:evaluations_mimo_ofdm}, we remove such assumptions and evaluate our method in more realistic 3GPP 3D channels.


\subsection{Effectiveness of PE layer}

We compare three approaches: 1) ADNN-GT: The ADNN with perfect channel knowledge; 2) ADNN-LMMSE: The ADNN with LMMSE estimated channel; 3) StructNet: our introduced approach.
The classifiers in all these three networks are comprised of two linear layers connected with the hyperbolic tangent non-linear function, and trained with the cross-entropy loss.
In Fig.~\ref{figs:toy_example_ser} (a), we plot the symbol error rate (SER) as a function of bit energy to noise ratio ($E_b/N_o$).
Compared with ADNN-LMMSE, StructNet can achieve better performance.
The performance gain is more significant with a relatively high $E_b/N_o$.
In addition, StructNet is shown to have comparable SER performance with ADNN-GT, where the perfect channel knowledge is used.
The results indicate that even if the PE layer starts from an inaccurate initialization, StructNet can achieve comparable performance with the approach exploiting the ground truth CSI.



\subsection{Experiments of training with incorrect labels}
\label{sec_exp:structnet_incorrect_labels}

In Sec.~\ref{sec:structnet_analysis_incorrect_labels}, we analyzed the reason why StructNet is robust to incorrect PAM labels.
In this section, we conduct the experiment to show how the performance of StructNet is affected by different percentages of incorrect training labels.
In the experiment, we randomly select a certain percentage of PAM samples to be incorrect.
The performance is evaluated under $5$ dB $E_b/N_o$.
Besides the two methods mentioned above, we also compare the performance with a four-class classifier.
For a fair comparison, the four-class classifier also adopts two linear layers and the hyperbolic tangent non-linear function, except that the output layer is of size $4$.
The results of SER versus the incorrect PAM label percentage are shown in Fig.~\ref{figs:toy_example_ser} (b).
As opposed to the general four-class classifier that is significantly affected by incorrect labels, StructNet performs reasonably well with even $70\%$ incorrect PAM labels.
This is because the $70\%$ incorrect PAM labels only contribute to $35\%$ incorrect binary labels, making it easier to handle by the network.
The results are consistent with our analysis and demonstrate the ability of StructNet to combat the incorrect training samples.
Such a property allows StructNet to mitigate the error propagation issue inherent in DF-based approaches, making it suitable to be applied with the DF mechanism.

\section{Evaluation with 3GPP-3D channel}
\label{sec:evaluations_mimo_ofdm}
In this section, we show the performance of RC-AttStructNet-DF both in the MIMO-OFDM system and the massive MIMO-OFDM system. 
Unlike the setting in RC-Struct~\cite{xu2021rcstruct}, the experiments are conducted at a user speed of $30$ km/h.
We compare the introduced approach with the conventional model-based methods and state-of-the-art learning-based strategies.


\begin{figure*}
\centering%
\vspace{-1em}
\subfloat[]{\label{a}\includegraphics[width=0.33\linewidth]{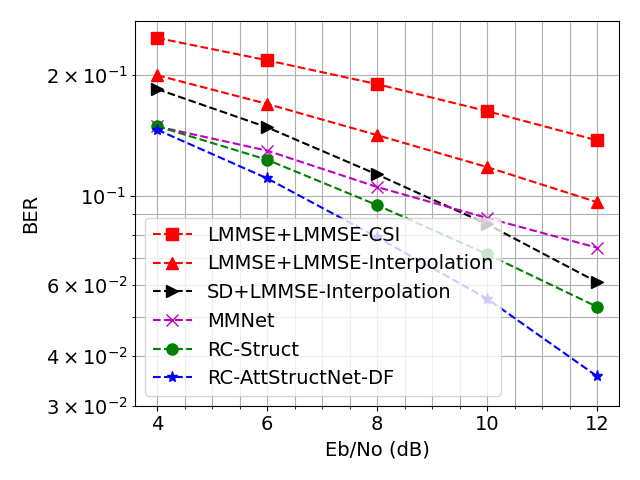}}%
\subfloat[]{\label{b}\includegraphics[width=0.33\linewidth]{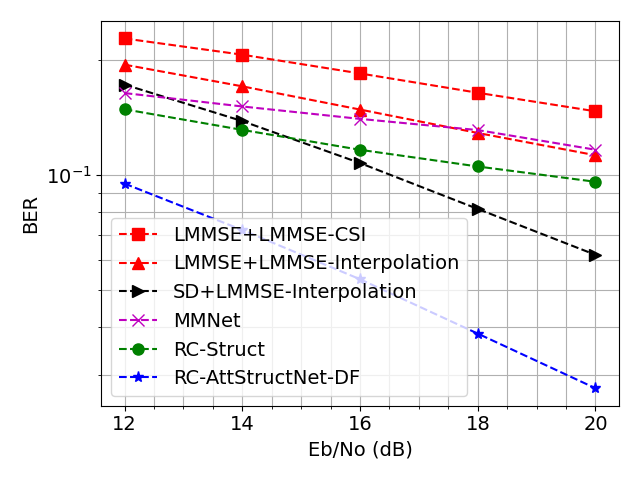}}
\subfloat[]{\label{c}\includegraphics[width=0.33\linewidth]{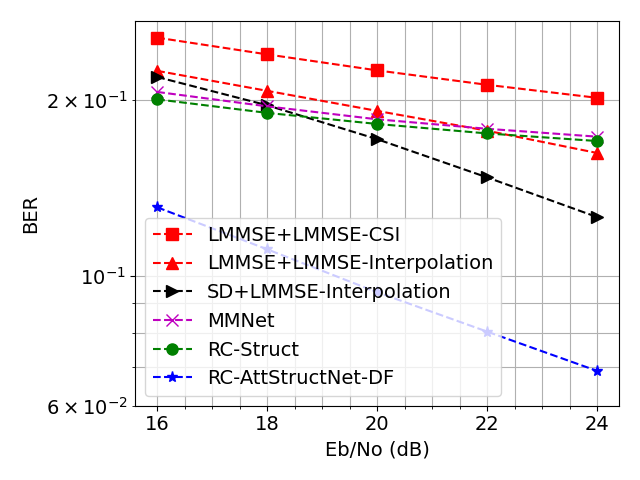}}
\caption{BER comparison in the MIMO-OFDM system.
(a) QPSK
(b) $16$ QAM
(c) $64$ QAM
\label{figs:mimo_linear}
}
\vspace{-1em}
\end{figure*}

\subsection{Experimental setting}
In the experiments, the number of subcarriers is set to be $N_{\mathrm{sc}} = 512$ and the CP length is $N_{\mathrm{cp}} = 32$.
Each subframe has a total of $N = 20$ OFDM symbols, among which $N_p = 4$ OFDM symbols are the training pilot and $N_d = 16$ OFDM symbols are the data symbols.
Note that only $4$ pilot symbols are used as the training data for each subframe, which is different from other learning-based approaches that exploit a large amount of training data. 
The wireless channels are generated following the 3GPP 3D MIMO model~\cite{study3d3gpp} with the QuaDRiGa simulator~\cite{jaeckel2014quadriga}.
The user speed is set as $30$ km/h, which is different from the setting in RC-Struct with a speed of $5$ km/h. 
In addition, gray coding is adopted in the experiments.

In terms of the setting of RC, the number of neurons is $N_n=16$, and the number of layers is $V=1$.
In~\cite{zhou2019}, it is shown that the utilization of a sliding window for the input to RC can increase the short-term memory capacity of RC.
Following this work, a sliding window of size $32$ is utilized to the input.
Note that all RC-based approaches compared in this work adopt the same sliding window for a fair comparison.
In the 2D MHA, the key embedding dimension for the time and frequency attention block is set as $216$ and $8$, respectively.
For the StructNet in the frequency domain, the input layer has $128$ neurons, and the output layer has $2$ neurons.
In StructNet, the offline weights of the binary classifier are trained with $2000$ randomly generated symbols for $1000$ epochs.
As choosing $E_b/N_o$ for training the offline weights does not provide any significant performance gain and is also not practical, the $E_b/N_o$ for each training sample is randomly chosen from $0$ dB to $15$dB to obtain relatively generic offline weights.
During online training, we adopt an alternative training strategy, which updates the PE layer and the binary classifier separately.
Specifically, we first update the binary classifier and fix the PE layer.
Then the PE layer is updated with the weights of the binary classifier fixed.
The total number of training epochs is set to be $20$.
In addition, to reduce the computation complexity, nine resource block groups (RBGs) are combined to train a single network. The probability threshold for attention loss is set as $\eta = 0.5$. It means that we select the detected symbols that the network predicts to have over $50\%$ chance to be correct. In other words, if the network thinks the detected data symbol has less than or equal to a $50\%$ of the chance being correct, the detected symbol is less likely to be correct and thus we do not use it in the DF procedure.

\subsection{BER Comparison in the MIMO-OFDM system and the massive MIMO-OFDM system}
We compare the following approaches:
(1) \emph{LMMSE+LMMSE-CSI}: The LMMSE-based symbol detector with LMMSE estimated CSI;
(2) \emph{LMMSE+LMMSE-Interpolation}: The LMMSE-based symbol detector using interpolated LMMSE CSI;
(3) \emph{SD+LMMSE-Interpolation}: The non-convex symbol detector that performs ML detection with SD approach utilizing interpolated LMMSE CSI~\cite{ghasemmehdi2011faster};
(4) \emph{MMNet}: The MMNet network proposed in~\cite{khani2020adaptive}, where the network for each subcarrier has been trained for $500$ iterations;
(5) \emph{RC-Struct}: The RC-based approach using LMMSE estimated shifting parameter in the frequency domain~\cite{xu2021rcstruct};
(6) \emph{RC-AttStructNet-DF}: The introduced method in this paper.
Note that the ``LMMSE-CSI" refers to that the LMMSE channel estimates only use the pilot symbols.
The ``LMMSE-Interpolation" means that the channel estimates are interpolated over data symbols using the pilot estimated CSI.


\begin{figure*}
\centering%
\vspace{-1em}
\subfloat[]{\label{a}\includegraphics[width=0.33\linewidth]{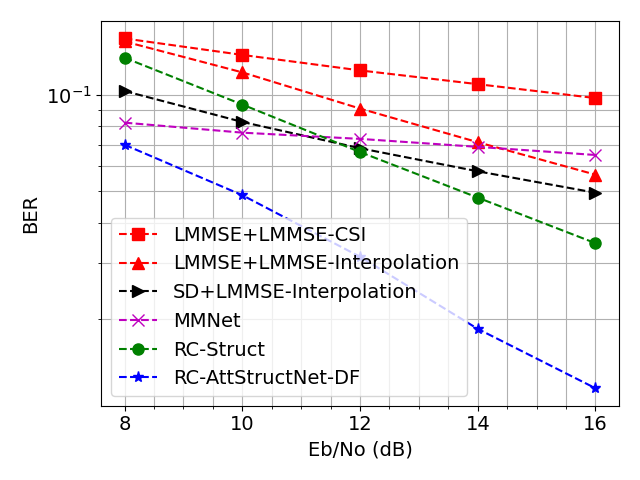}}%
\subfloat[]{\label{b}\includegraphics[width=0.33\linewidth]{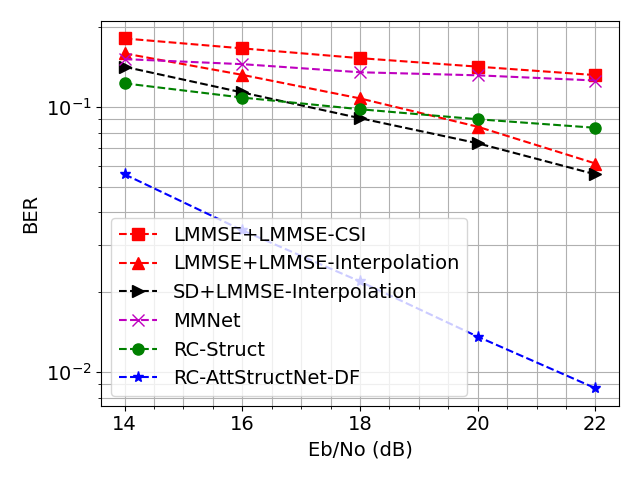}}
\subfloat[]{\label{c}\includegraphics[width=0.33\linewidth]{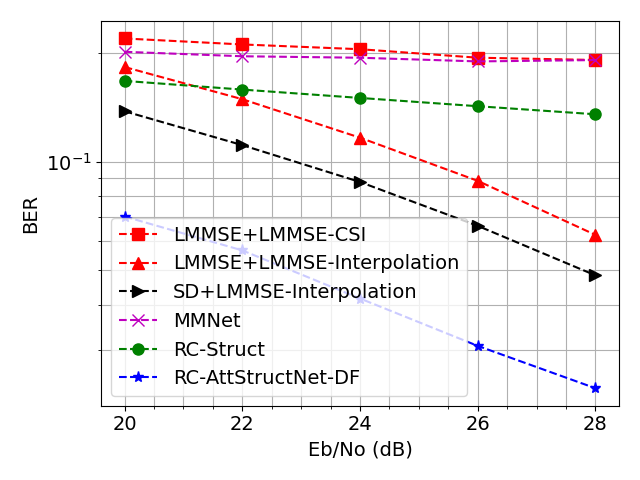}}
\caption{BER comparison in the massive MIMO-OFDM system.
(a) QPSK
(b) $16$ QAM
(c) $64$ QAM
\label{figs:massive_mimo_linear}
}
\vspace{-1em}
\end{figure*}

We conduct experiments in two system settings: the MIMO-OFDM system and the massive MIMO-OFDM system.
In the MIMO-OFDM system, the number of transmit antennas is set as $N_t = 4$ and the number of receive antennas is $N_r = 4$.
In Fig.~\ref{figs:mimo_linear}, we show the BER plot for QPSK, $16$ QAM, $64$ QAM, respectively.\footnote{Note that the BER of RC-AttStructNet-DF (without channel coding) ranges from $3\%$ to $14\%$, which lies within the typical BER range specified by the 3GPP 5G NR~\cite{std3gpp38214, std3gpp38133}.
For instance, the user equipment (UE) channel quality indicator (CQI) calculation is based on a target block error rate (BLER) of 10\% (after channel coding)~\cite{std3gpp38214}, and radio link monitoring out-of-sync BLER is set to be 10\%~\cite{std3gpp38133}.}
The results show that all the learning-based approaches and the SD method outperform LMMSE when the CSI is obtained by using only the pilot symbols for all the tested modulation orders.
With the interpolated CSI, the BER of the LMMSE detection scheme decreases, as the channel estimates become more accurate when interpolated over the data symbols.
As exhibited in Fig.~\ref{figs:mimo_linear} (b) and Fig.~\ref{figs:mimo_linear} (c), the SD approach has better performance than the RC-Struct and MMNet in the high $Eb/No$ regime when $16$ QAM and $64$ QAM modulation orders are used.
However, in the low $Eb/No$ regime, the performance of SD becomes worse than RC-Struct and MMNet due to inaccurate channel estimates.
The reason is that the channel estimates in the low $Eb/No$ regime are more precise than in the high $Eb/No$ regime, leading to performance degradation.
These observations indicate that the performance of the conventional approaches LMMSE and SD heavily relies on the accuracy of the CSI estimation.
The inherent error of the CSI estimation can jeopardize the detection performance.
On the other hand, as a learning-based approach that does not rely on explicit channel estimation, RC-AttStructNet-DF is not affected by such impairments and is shown to have outstanding performance gain over conventional methods for all the tested modulation orders.

Regarding the learning-based approaches, in Fig.~\ref{figs:mimo_linear}, we can see that the RC-AttStructNet-DF consistently outperforms the MMNet algorithm and the RC-Struct method under different scenarios.
The reason is that for MMNet, it requires a larger amount of training data than the setup in this paper to learn the network weights.
As an online over-the-air scenario is adopted in our evaluation, the MMNet learned by the limited amount of training data suffers from the model overfitting problem, resulting in performance degradation.
Different from MMNet, by embedding the structural knowledge of the MIMO-OFDM system, RC-AttStructNet-DF can be learned with limited training data in an online fashion.
Furthermore, both the RC-Struct and the MMNet only learn from the pilot symbols.
Due to the relatively high user mobility, the neural network weights only trained by the pilot symbols are not sufficient to track the changes of the channel over data symbols, and thus have an unsatisfactory detection performance when testing on the data symbols.
Instead, RC-AttStructNet-DF dynamically updates the network weights according to the channel variation with the data symbols.
With the specially designed architecture and the dynamic adaptation, better performance is achieved by the RC-AttStructNet-DF.

In the massive MIMO-OFDM system, we test in an uplink scenario with 4 transmit antennas and 64 receive antennas. 
In particular, at the transmitter side, the number of scheduled UE is 2, where each UE has 2 transmit antennas.
At the receiver side, the base station (BS) is equipped with a rectangular planar array that has 8 azimuth antennas and 8 elevation antennas.
Fig.~\ref{figs:massive_mimo_linear} shows the BER performance in the massive MIMO-OFDM system with QPSK, $16$ QAM, and $64$ QAM.
The same trend holds as in the MIMO-OFDM system, where RC-AttStructNet-DF achieves the lowest BER.
The results further demonstrate the advantages of RC-AttStructNet-DF over the other methods under different scenarios.

\subsection{BER comparison with nonlinear distortion}



\begin{figure}
\centering
\includegraphics[width=0.8\linewidth]{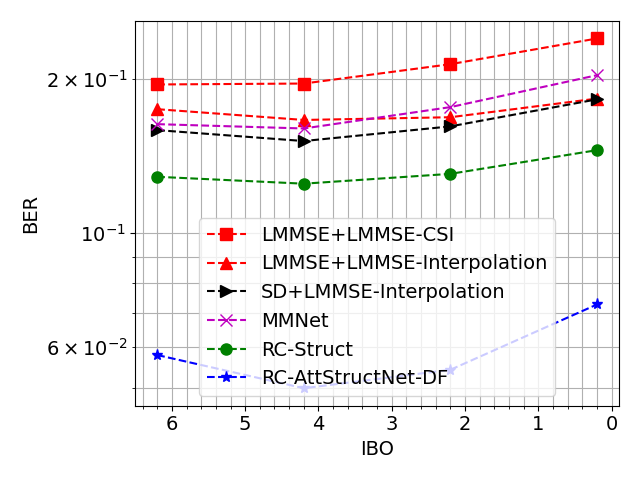}
\caption{BER comparison in the nonlinear region.}
\label{figs:ber_mimo_16qam_nonlinear}
\vspace{-1em}
\end{figure}

\begin{figure*}
\centering%
\subfloat[]{\label{a}\vspace{-1em}\includegraphics[width=0.8\linewidth]{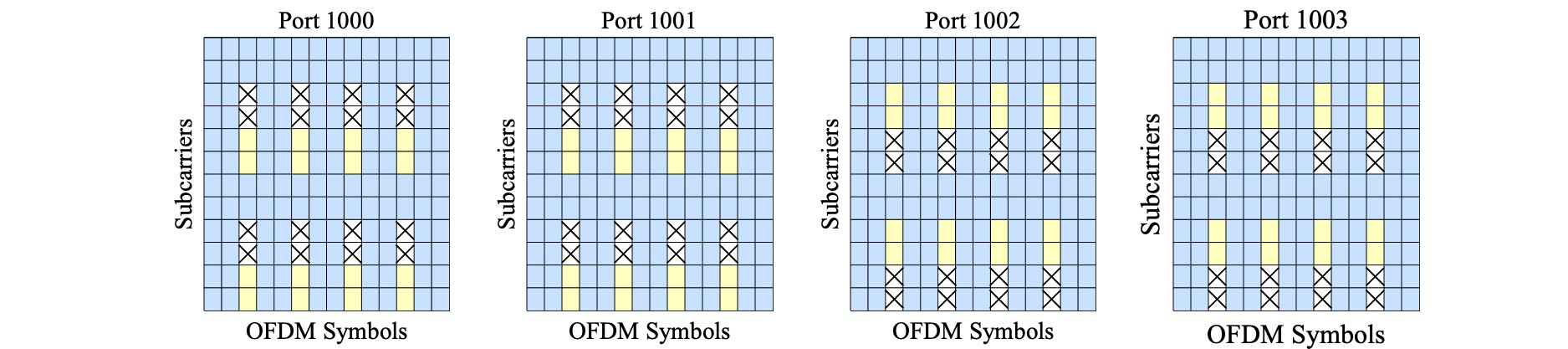}}%
\\
\vspace{-1em}
\subfloat[]{\label{b}\vspace{-1em}\includegraphics[width=0.8\linewidth]{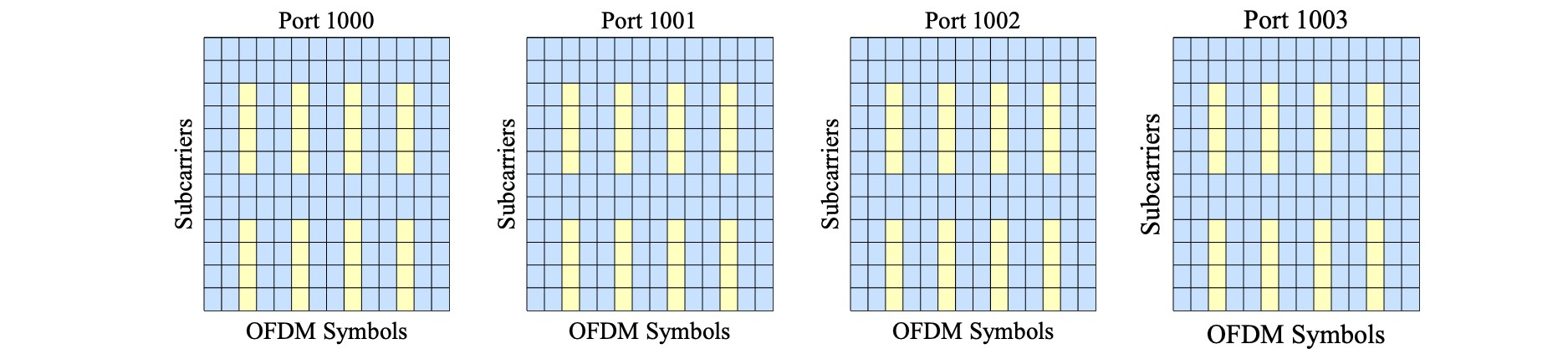}}
\caption{The MIMO scattered pilot pattern in one resource block (RB).
(a) Conventional approaches.
(b) RC-AttStructNet-DF.
\label{figs:mimo_scattered_pilot_pattern}
}
\vspace{-1em}
\end{figure*}

In this section, we perform the experiment when PA distortion is applied to the input signal.
The PA model $g(x) = \frac{x}{\left[1 + \left(\frac{|x|}{x_{\mathrm{sat}}}\right)^{2\rho}\right]^{0.5\rho}}$ is adopted to introduce channel distortion~\cite{rapp1991effects}, where $x$ is the input transmitted signal, $x_{\mathrm{sat}}$ is the PA saturation level, and $\rho$ measures the smoothing parameter.
We define the input back-off (IBO) as the ratio between PA’s saturation power to the input power.
The input signal is distorted when the peak-to-average-power ratio (PAPR) of the input signal is higher than IBO. 
We adopt $x_{\mathrm{sat}} = 1$ and $\rho = 3$ and set the non-linear region as the case when IBO is smaller than $6.5$ dB.
The experiments are conducted in the $4\times 4$ MIMO-OFDM system with $16$ QAM modulation.
Fig.~\ref{figs:ber_mimo_16qam_nonlinear} shows the performance when the non-linear distortion exists.
The performance of the conventional schemes, such as the SD and LMMSE, is highly affected by the system's nonlinearity.
Specifically, the BER of the SD method increases quickly when the IBO reduces, as the estimated CSI becomes less accurate with stronger signal distortion.
As the distortion level increases, the performance of the MMNet approach also degrades due to the linear assumption for its system model.
Furthermore, RC-based approaches perform better than the conventional approaches when IBO is low, indicating that the RC-based approaches are better at combating the nonlinearity.
More importantly, RC-AttStructNet-DF achieves the best performance among all the schemes, demonstrating its generalization ability in different cases.






\subsection{Effectiveness of attention mechanism, PE layer, and DF}

\begin{figure}
\centering
\includegraphics[width=0.8\linewidth]
{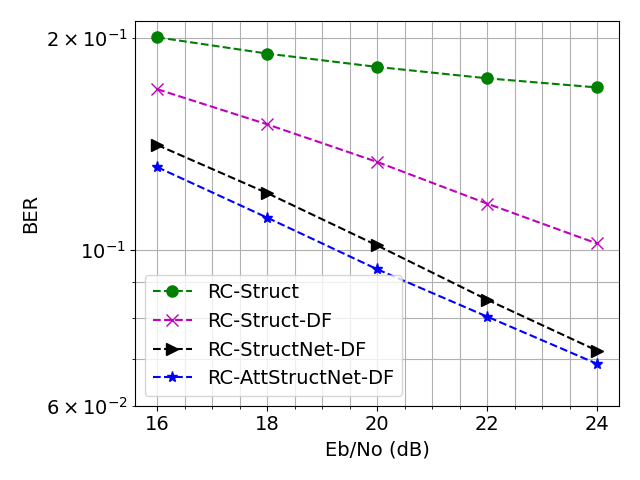}
\caption{BER for testing effectiveness of different modules.}
\vspace{-1em}
\label{figs:ber_mimo_effect_attention_pe_layer_64qam}
\end{figure}

In this section, we demonstrate the effectiveness of incorporating the attention mechanism, PE layer, and DF.
The experiments are conducted in the MIMO-OFDM system with $4$ transmit antennas and $4$ receive antennas.
The modulation order is set as $64$ QAM.
In Fig.~\ref{figs:ber_mimo_effect_attention_pe_layer_64qam}, we compare the BER performance of four methods: (1) RC-Struct; (2) RC-Struct-DF: RC-Struct with DF; (3) RC-StructNet-DF: the approach with PE layer and DF; (5) RC-AttStructNet-DF: our introduced approach with the attention mechanism, PE layer, and DF.
As shown in Fig.~\ref{figs:ber_mimo_effect_attention_pe_layer_64qam}, RC-Struct-DF outperforms RC-Struct, which indicates the effectiveness of using DF when structural information is incorporated in the network.
By comparing the performance of RC-StructNet-DF with RC-Struct-DF, we can see that the BER performance is further boosted when the PE layer is adopted in the frequency domain.
The results suggest that the introduced PE layer in StructNet can further facilitate the DF mechanism and improve the detection performance by allowing the network to dynamically update the network parameters according to channel variations.
Moreover, RC-AttStructNet-DF achieves an additional performance gain over RC-StructNet-DF, demonstrating that the attention mechanism is a valuable addition to our network.


\subsection{BER comparison with practical pilot pattern}
\label{exp:scattered_pilot_pattern_exp}

While our previous discussions are all based on the pilot structure where the first $N_p$ OFDM symbols are all pilots, in this section, we show that our introduced approach can also be applied to a scattered pilot pattern specified by the 3GPP standard.
We consider the MIMO pilot pattern shown in Fig.~\ref{figs:mimo_scattered_pilot_pattern} following the 3GPP 5G NR~\cite{std3gpp36211, std3gpp36212}. 
The pilot resource elements (REs) are colored in yellow and the data REs are colored in blue. 
The white RE represents the empty pilot symbols. 
As illustrated in Fig.~\ref{figs:mimo_scattered_pilot_pattern} (a), for conventional approaches, empty pilot symbols are transmitted.
In addition, pilots are set to be orthogonal among different antenna ports.
The empty and orthogonal settings of the pilot are utilized to eliminate pilot interference and ensure a more accurate MIMO channel estimation.
For our introduced learning-based method, we adopt the pilot structure in Fig.~\ref{figs:mimo_scattered_pilot_pattern} (b), where all the pilots are randomly generated. 
It is noteworthy that the pilot pattern in Fig.~\ref{figs:mimo_scattered_pilot_pattern} (b) has the same training overhead as the pilot structure in Fig.~\ref{figs:mimo_scattered_pilot_pattern} (a). 
The difference is that we try to avoid pilot interference using the empty and orthogonal pilots for conventional approaches, while we keep the pilot interference with the pilot pattern for learning-based methods. 
The reason is that, for conventional detectors that rely on channel estimation, the received pilots should be free of interference to obtain a more accurate estimated CSI. 
However, for learning-based methods, the neural network needs to learn the situation when interference exists to avoid the mismatch between the training stage and the testing stage. 
More detailed discussions about the pilot pattern design have been provided in our previous work~\cite{zhou2019}.

\begin{figure}
\centering
\includegraphics[width=0.8\linewidth]{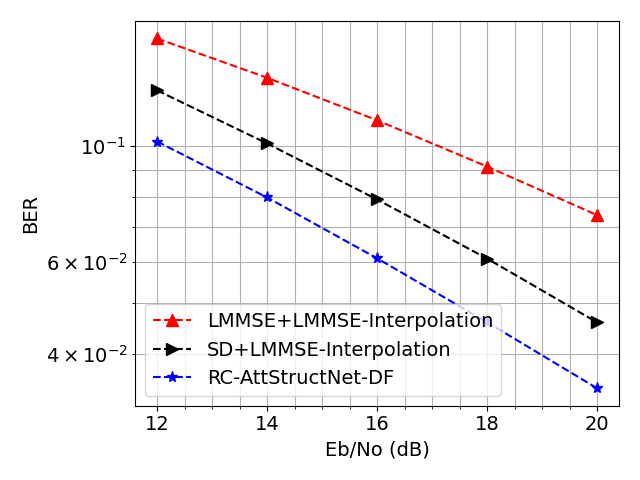}
\caption{BER comparison with the scattered pilot pattern.}
\label{figs:ber_mimo_scattered_pilot}
\vspace{-1em}
\end{figure}

The experiment is conducted in the $4\times4$ MIMO-OFDM system with 16 QAM modulation. 
The total number of OFDM symbols is set as 14 following the 3GPP 5G NR standard~\cite{std3gpp36211, std3gpp36212}.
The training overhead of this scattered pilot pattern is approximately $19\%$.
Note that the block pilot pattern, where there are $20$ OFDM symbols in total and the first $4$ OFDM symbols are pilots, has a training overhead of $20\%$, which also satisfies the pilot occupancy requirement specified in~\cite{std3gpp36211, std3gpp36212}.
With the scattered pilot pattern, the RC-AttStructNet-DF network is first trained with the pilot REs to obtain the initial weights. Then the DF is conducted for the full subframe, including the OFDM symbol with pilot REs, in a symbol-by-symbol manner.
During the DF procedure, when re-training with the detected OFDM symbol that has pilot REs, the pilot positions are filled up with the ground truth pilots instead of detected pilots to generate training labels.
As displayed in Fig.~\ref{figs:ber_mimo_scattered_pilot}, our introduced RC-AttStructNet-DF approach outperforms both the conventional LMMSE and SD methods with this practical scattered pilot pattern. 
More importantly, the results further demonstrate the generalization ability of RC-AttStructNet-DF and its potential to be adopted in a realistic setting.

\subsection{BER comparison with conventional methods using decision-directed channel estimation}
\begin{figure}[htp]
    \centering
    \begin{minipage}{0.45\textwidth}
        \centering
        \includegraphics[width=0.8\textwidth]{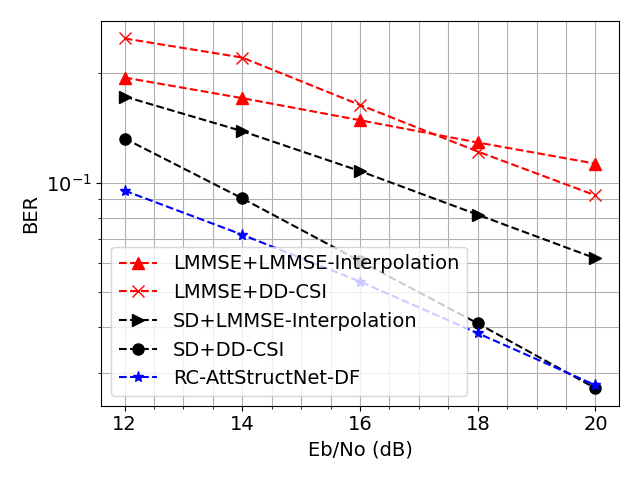}
        \caption{BER comparison with conventional schemes using DD-CSI.}
        \label{figs:mimo_decision_directed_linear}
    \end{minipage}
    \\
    \vspace{1em}
    \begin{minipage}{0.45\textwidth}
        \centering
        \includegraphics[width=0.8\textwidth]{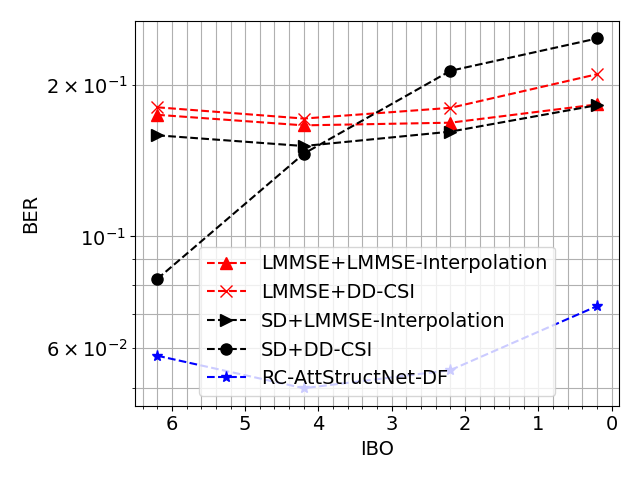}
        \caption{BER comparison with adopting DD-CSI in nonlinear region.}
        \label{figs:mimo_decision_directed_nonlinear}
    \end{minipage}
    \vspace{-1em}
\end{figure}

In this section, we conduct the performance comparison with conventional approaches when the RLS-based decision-directed (DD) channel estimation~\cite{karami2006decision} is adopted.
Specifically, the first four pilot OFDM symbols are utilized to obtain the initial channel estimation using the RLS scheme.
Then the subsequent CSI corresponding to the data symbols is obtained by treating the detected symbols as training data and estimated iteratively using the RLS approach.
The BER performance is provided in Fig.~\ref{figs:mimo_decision_directed_linear}. 
The LMMSE detector and SD detector with this DD channel estimation are referred to as ``LMMSE+DD-CSI” and ``SD+DD-CSI”, respectively. 
The experiment is performed under the $4\times4$ MIMO-OFDM system with $16$ QAM modulation. 
As shown in Fig.~\ref{figs:mimo_decision_directed_linear}, by dynamically estimating the channel of data symbols, ``LMMSE+DD-CSI” has better performance than the LMMSE detector with interpolated LMMSE CSI in the high $E_b/N_o$ regime. 
However, due to the error propagation caused by the inaccurate data symbol detection, the performance of ``LMMSE+DD-CSI” is degraded in mid to low $E_b/N_o$ regimes. 
The performance of the SD detector is boosted by the DD channel estimates when compared with the SD detector utilizing the interpolated LMMSE CSI.
Compared with conventional approaches, our introduced RC-AttStructNet-DF achieves better performance than ``LMMSE+DD-CSI” across different $E_b/N_o$’s and outperforms ``SD+DD-CSI” in the mid to low $E_b/N_o$ regimes.
We further compare the performance when the PA distortion is applied to the input signal. 
As indicated in Fig.~\ref{figs:mimo_decision_directed_nonlinear}, the performance of the ``LMMSE+DD-CSI” and ``SD+DD-CSI” deteriorate when signal distortion becomes stronger, especially for the “SD+DD-CSI”.
On the other hand, RC-AttStructNet-DF is less affected by the nonlinearity within the system and exhibits better performance than the conventional approaches.

\subsection{Empirical complexity of symbol detection approaches}

To see the performance-complexity  trade-off more straightforwardly, we compare the empirical complexity of different methods under the same setting for Fig.~\ref{figs:mimo_linear} (c). 
Specifically, we show CPU run time for different methods in the setting of the MIMO-OFDM system with 4 transmit antennas and 4 receive antennas and 64 QAM modulation. 
In Tab.~\ref{tab:cpu_run_time}, we present both the training time and the total processing time for a subframe of different methods, where the total processing time includes both the training and testing time for a subframe.
Tab.~\ref{tab:cpu_run_time} indicates that RC-AttStructNet-DF has a longer total processing time than LMMSE-based approaches. 
However, RC-AttStructNet-DF can achieve a 65\% and 57\% BER reduction over the LMMSE with LMMSE-CSI and LMMSE with LMMSE-Interpolation under 24 dB $E_b/N_o$, respectively, as shown in Fig.~\ref{figs:mimo_linear} (c). 
While RC-AttStructNet-DF takes a longer total processing time than RC-Struct, the CPU run time of these two approaches are still on the same scale. 
With the same scale of the total processing time, RC-AttStructNet-DF performs significantly better than RC-Struct. 
On the other hand, when compared with the SD method, RC-AttStructNet-DF processes in a shorter amount of time and is also demonstrated to have a better performance. 
More importantly, when compared with the state-of-the-art learning-based approach MMNet, RC-AttStructNet-DF processes over 100 times faster, which demonstrates the potential of RC-AttStructNet-DF to be adopted for online detection.

\begin{table}
\centering
\caption{CPU run time of symbol detection methods}
\resizebox{\linewidth}{!}{
\begin{tabular}{lcc}
\toprule
Method & Training time (Sec.) & Total processing time (Sec.) \\
\midrule
LMMSE+LMMSE-CSI & - & 0.12 \\
LMMSE+LMMSE-Interpolation & - & 7.57 \\
SD+LMMSE-Interpolation & - & 152.06 \\
RC-Struct & 17.55 & 18.27 \\
RC-AttStructNet-DF & 26.67 & 29.57 \\
MMNet & 3014.52 &  3025.66 \\
\bottomrule
\end{tabular}
}
\label{tab:cpu_run_time}
\end{table}

\section{Conclusion}
\label{sec:conclusion}
In this paper, we introduce an online attention-based approach, RC-AttStructNet-DF, for conducting MIMO-OFDM symbol detection on a subframe basis.
The approach employs reservoir computing (RC) in the time domain.
The frequency domain network consists of the 2D MHA module along with a structure-based network StructNet, which is learned with an attention loss.
The 2D MHA is exploited to capture the time and frequency correlations of the signal.
The StructNet is designed to mitigate the error propagation of the DF approach and facilitate the DF mechanism. 
With the StructNet, the adopted DF mechanism further enhances detection performance by learning from detected data symbols and dynamically tracking channel changes within a subframe.
Extensive experiments in 3GPP 3D channels demonstrate the effectiveness of RC-AttStructNet-DF in detection under different scenarios with the BER performance and the effectiveness of different modules.
\ifCLASSOPTIONcaptionsoff
  \newpage
\fi



\bibliographystyle{IEEEtran}

\bibliography{IEEEabrv,ref.bib}
%

%







\end{document}